\documentclass[camera,letterpaper,nomarginnotes,nonarrowgutter,hyperref]{jpaper}

\usepackage{amsmath,amssymb,amsfonts}
\usepackage{algorithmic}
\usepackage{graphicx}
\usepackage{textcomp}
\usepackage{xcolor}

\usepackage{subcaption}
\usepackage{enumitem}
\usepackage{tabularx}
\usepackage{booktabs}
\usepackage[dvipsnames]{xcolor}
\usepackage{makecell}
\usepackage{wasysym}

\usepackage[table]{xcolor}
\usepackage{needspace}
\usepackage{setspace}
\usepackage[sort,compress]{cite}

\usepackage{xcolor, soul}
\usepackage{xspace}
\usepackage{multirow}
\usepackage{tabularx}
\usepackage{tikz}
\usepackage[bottom]{footmisc}
\usepackage[most]{tcolorbox}

\usepackage{fancyhdr}
\usepackage{mathtools}

\usepackage{subcaption}

\usepackage{booktabs}

\usepackage{pgfplots}
\pgfplotsset{compat=1.17} 

\usepackage{stfloats}

\usepackage{enumitem}

\usepackage{mwe}

\usepackage[most]{tcolorbox}

\definecolor{lavenderr}{rgb}{0.71, 0.49, 0.86}

\usepackage[most]{tcolorbox}   
\tcbset{
  observebox/.style={
    colframe=lavenderr!90!black,      
    colback=lavenderr!15,              
    boxrule=1pt,                   
    arc=1mm,                       
    left=0.1mm, right=0.1mm,       
    top=0.1mm, bottom=0.1mm,       
    fonttitle=\bfseries,           
    before skip=6pt, after skip=6pt
  }
}

\usepackage{graphicx}

\usepackage{tikzducks}

\usepackage[en-GB]{datetime2} 
\DTMsetup{
    showzone=false,   
    showseconds=false 
}

\definecolor{darkspringgreen}{rgb}{0.09, 0.45, 0.27}
\definecolor{denim}{rgb}{0.08, 0.38, 0.74}
\definecolor{darkolivegreen}{rgb}{0.33, 0.42, 0.18}
\definecolor{tangerine}{rgb}{0.95, 0.52, 0.0}
\definecolor{mahogany}{rgb}{0.75, 0.25, 0.0}
\definecolor{coolblack}{rgb}{0.0, 0.22, 0.44}
\definecolor{darkpink}{rgb}{0.91, 0.35, 0.6}
\definecolor{darkblue}{rgb}{0.0, 0.0, 0.67}
\definecolor{melon}{rgb}{0.97, 0.69, 0.67}

\definecolor{seagreen}{rgb}{0.18, 0.55, 0.34}
\definecolor{pred}{rgb}{0.7843, 0.0039, 0.3137} 

\definecolor{darkpink}{rgb}{0.88, 0.28, 0.54}
\definecolor{forestgreen}{rgb}{0.0, 0.27, 0.13}
\definecolor{amber}{rgb}{1.0, 0.49, 0.0}

\newcolumntype{Y}{>{\centering\arraybackslash}X}
\usepackage{tikz}

\newcommand{\squishlist}{
 \begin{list}{$\circ$}
  { \setlength{\itemsep}{0pt}
     \setlength{\parsep}{0pt}
     \setlength{\topsep}{3pt}
     \setlength{\partopsep}{0pt}
     \setlength{\leftmargin}{1em}
     \setlength{\labelwidth}{1em}
     \setlength{\labelsep}{0.5em} } }

\newcommand{\squishend}{
  \end{list}  }

\makeatletter
\g@addto@macro{\normalsize}{%
  \setlength{\abovedisplayskip}{4pt plus 0.5pt minus 1pt}
  \setlength{\belowdisplayskip}{4pt plus 0.5pt minus 1pt}
  \setlength{\abovedisplayshortskip}{0pt}
  \setlength{\belowdisplayshortskip}{0pt}
  \setlength{\intextsep}{3pt plus 1pt minus 1pt}
  \setlength{\textfloatsep}{7pt plus 1pt minus 1pt}
  \setlength{\skip\footins}{4pt plus 1pt minus 1pt}}
\makeatother

\usepackage{blindtext,graphicx}
\usepackage[absolute]{textpos}
\usepackage{datetime}

\definecolor{seagreen}{rgb}{0.18, 0.55, 0.34}
\definecolor{ballblue}{rgb}{0.13, 0.67, 0.8}

\definecolor{darkgreen}{rgb}{0.0, 0.44, 0.34}

\sloppypar

\definecolor{dollarbill}{rgb}{0.52, 0.73, 0.4}

\usetikzlibrary{patterns}
\usepackage[precision=2, unit=mm]{lengthconvert}

\usepackage[colorinlistoftodos,prependcaption,textsize=scriptsize]{todonotes}
\usepackage{marginnote} 
\definecolor{cyan(process)}{rgb}{0.0, 0.62, 0.82}

\definecolor{cadmiumgreen}{rgb}{0.0, 0.50, 0.29}

\newcommand\revref[1]{\hyperref[rev:#1]{#1}}

\definecolor{raspberry}{rgb}{0.89, 0.04, 0.36}

\definecolor{awesome}{rgb}{1.0, 0.13, 0.32}
\definecolor{cardinal}{rgb}{0.77, 0.12, 0.23}
\definecolor{cadet}{rgb}{0.33, 0.41, 0.47}
\definecolor{celadon}{rgb}{0.67, 0.88, 0.69}
\definecolor{persianblue}{rgb}{0.11, 0.22, 0.73}
\definecolor{ultramarine}{rgb}{0.07, 0.04, 0.56}
\definecolor{warmblack}{rgb}{0.0, 0.3, 0.3}

\definecolor{terracotta}{rgb}{0.89, 0.45, 0.36}

\definecolor{forestgreen(web)}{rgb}{0.13, 0.55, 0.13}

\definecolor{cardinal}{rgb}{0.77, 0.12, 0.23}
\definecolor{deeppink}{rgb}{1.0, 0.08, 0.58}
\definecolor{brightpink}{rgb}{1.0, 0.0, 0.5}
\definecolor{electricviolet}{rgb}{0.56, 0.0, 1.0}
\definecolor{brandeisblue}{rgb}{0.0, 0.44, 1.0}
\definecolor{carminered}{rgb}{1.0, 0.0, 0.22}

\usepackage{setspace}
\definecolor{acolor}{rgb}{0.0, 0.5, 1.0}
\definecolor{bcolor}{rgb}{0.54, 0.17, 0.89}
\definecolor{ccolor}{rgb}{0.4, 0.69, 0.2}
\definecolor{dcolor}{rgb}{0.92, 0.41, 0.12}
\definecolor{ecolor}{rgb}{0.6, 0.0, 0.156}
\definecolor{fcolor}{rgb}{0.106, 0.620, 0.467}

\definecolor{dogwoodrose}{rgb}{0.84, 0.09, 0.41}

\usepackage{tcolorbox}
\usepackage{xspace}
\newcommand{\SysName}{\texttt{DCC}\xspace} 

\newcommand\cg[1]{\noindent{\color{black} {#1}}} %
\newcommand\pyang[1]{\noindent{\color{blue} {#1}}} %

\newcommand\revA[1]{\noindent{\color{black} {#1}}} %
\newcommand\revB[1]{\noindent{\color{black} {#1}}} %
\newcommand\revC[1]{\noindent{\color{black} {#1}}} %
\newcommand\revD[1]{\noindent{\color{black} {#1}}} %
\newcommand\revE[1]{\noindent{\color{black} {#1}}} %
\newcommand\revF[1]{\noindent{\color{black} {#1}}} %
\newcommand\revS[1]{\noindent{\color{black} {#1}}} %
\newcommand\CamOne[1]{\noindent{\color{black} {#1}}} %
\newcommand\ArXiv[1]{\noindent{\color{black} {#1}}} %

\newcommand{\kernelmaxattaccgpuspeedup}{13.17}
\newcommand{\kernelmaxhbmpimgpuspeedup}{7.68}
\newcommand{\inferenceMaxattaccgpuspeedup}{7.71}
\newcommand{\inferenceMaxattaccBaselinespeedup}{2.74}

\newcommand{\kernelavgattaccgpuspeedup}{\revC{3.92}}
\newcommand{\kernelavghbmpimgpuspeedup}{\revC{2.21}}
\newcommand{\inferenceAvgattaccgpuspeedup}{\revC{4.52}}
\newcommand{\inferenceAvgattaccBaselinespeedup}{\revC{1.75}}

\newcommand{\kerneltrainingtimeshort}{$\sim$67\xspace}

\def\BibTeX{{\rm B\kern-.05em{\sc i\kern-.025em b}\kern-.08em
    T\kern-.1667em\lower.7ex\hbox{E}\kern-.125emX}}

\hypersetup{
    colorlinks=true,
    urlcolor=blue,
}

\begin{document}

\title{\SysName: \textbf{D}ata-\textbf{C}entric \textbf{C}ompilation of Machine Learning Kernels \\ for Processing-In-Memory Architectures\vspace{-4pt}}
\def\iscacameraready{} 

\newcommand\iscaauthors{
Peiming Yang$^\ddagger$ \hspace{0.5em} Sankeerth Durvasula$^\ddagger$\hspace{0.5em} Ivan Fernandez$^\P$\hspace{0.5em} Mohammad Sadrosadati$^\S$ \\
Onur Mutlu$^\S$\hspace{0.5em} Gennady Pekhimenko$^{\ddagger,*}$\hspace{0.5em} Christina Giannoula$^\dagger$ 
{\vspace{2pt}\textcolor{white}{}}
\\
}

\newcommand\iscaaffiliation{

\emph{
\normalsize 
$^\ddagger$University of Toronto \& Vector Institute \hspace{0.5em}
$^\P$Barcelona Supercomputing Center
\vspace{-1pt}
} \\
\emph{
\normalsize 
$^\S$ETH Zürich \hspace{0.5em}
$^*$Nvidia \hspace{0.5em}
$^\dagger$Max Planck Institute for Software Systems}
}

\author{
\iscaauthors{}
\iscaaffiliation{}
\vspace{-10pt}
}

\pagestyle{fancy}
\fancyhf{} 
\renewcommand{\headrulewidth}{0pt}



\maketitle

\begin{tikzpicture}[remember picture, overlay]
\node[anchor=north east, inner sep=0pt, xshift=-0.25in, yshift=-0.11in] at (current page.north east) {%
\includegraphics[height=0.55in]{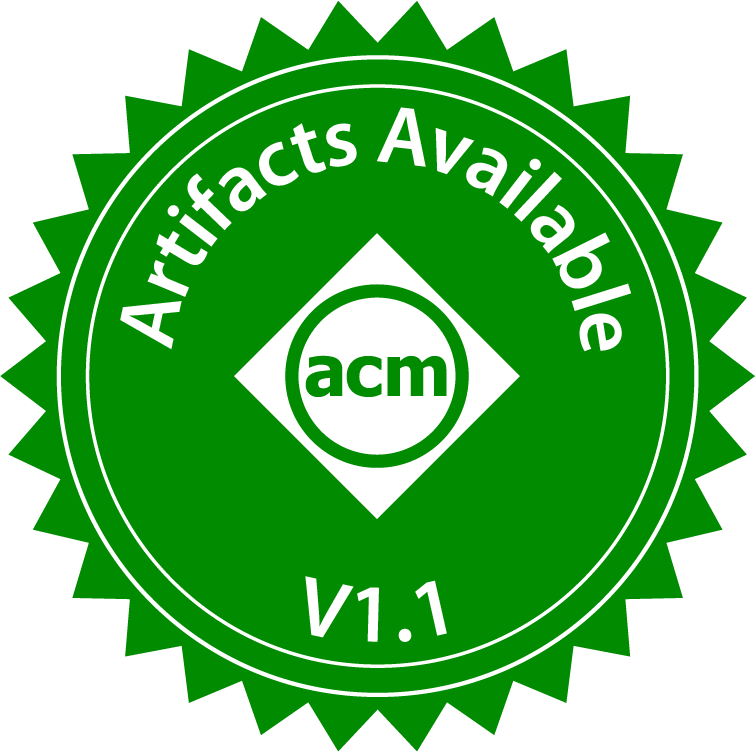}\hspace{0.06in}%
\includegraphics[height=0.55in]{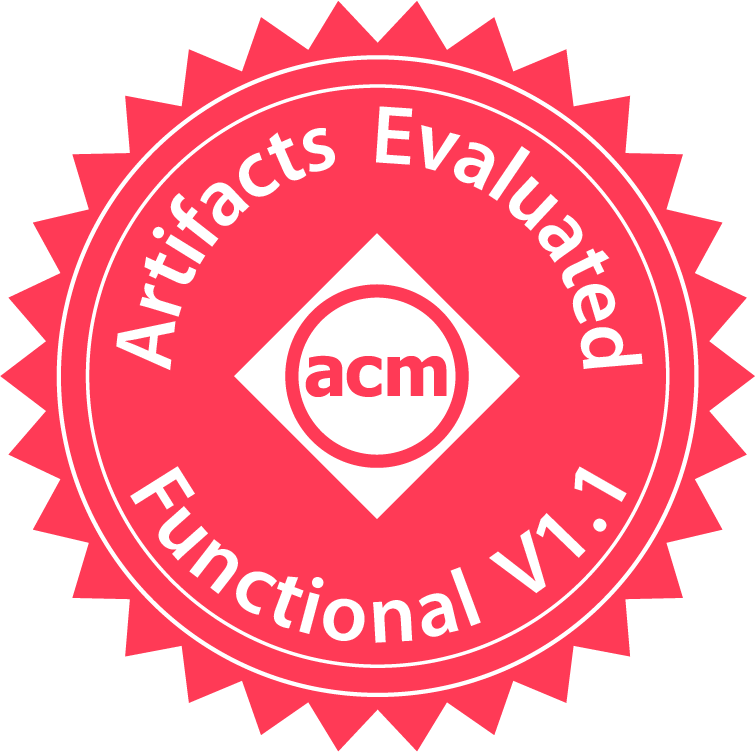}\hspace{0.06in}%
\includegraphics[height=0.55in]{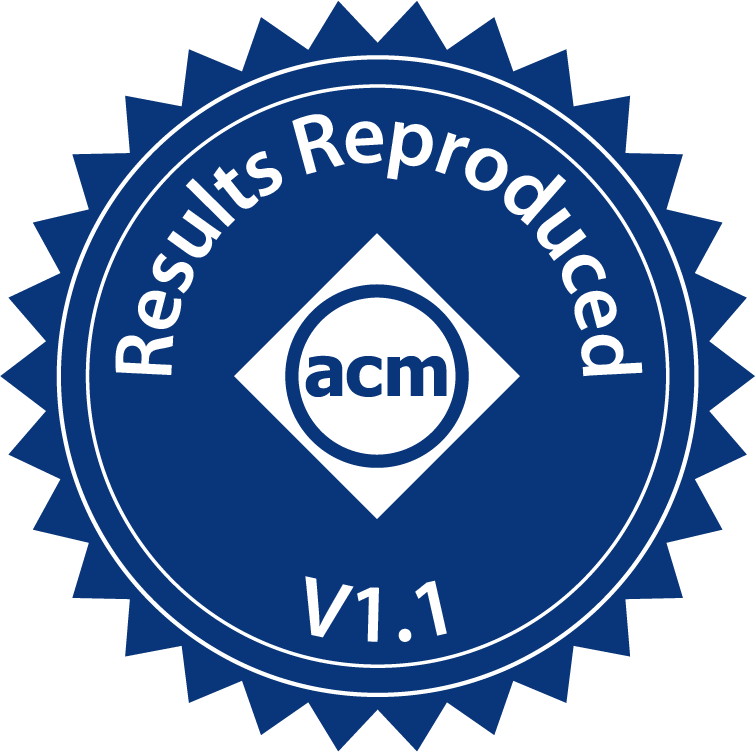}%
};
\end{tikzpicture}%

\newcommand{\iscaheight}{0mm}
\ifdefined\eaopen
\renewcommand{\iscaheight}{12mm}
\fi

\setcounter{page}{1}

\begin{abstract}
High-performance Host processors (e.g., GPUs) \CamOne{can} integrate Processing-In-Memory (PIM) devices, which  can accelerate memory-intensive kernels of Machine Learning (ML) models, including Large Language Models (LLMs), by leveraging the large memory bandwidth available at PIM cores. However, Host processor and PIM cores require different data layouts: Host processor needs consecutive elements distributed \emph{across} DRAM banks, while PIM cores need consecutive elements \emph{within} their local banks. This necessitates data rearrangements in ML kernel execution that pose significant performance and programmability challenges, further exacerbated by the need to support diverse PIM \CamOne{devices (e.g., Samsung HBM-PIM, SK Hynix GDDR6-AiM)}. 
Current compilation approaches lack systematic optimization for diverse ML kernels and multiple PIM \CamOne{devices}, and may largely ignore data rearrangement \CamOne{costs} during the compute code optimization \CamOne{step}. 
We demonstrate that data rearrangements and compute code optimization are interdependent, and need to be jointly optimized during the tuning process.
To address this, we design \SysName, the first data-centric ML compiler for PIM systems that jointly co-optimizes data rearrangements and compute code in a unified tuning process \CamOne{to enable high performance execution}. \SysName integrates a multi-layer PIM abstraction that enables various data distribution strategies on different PIM backends. \SysName enables effective co-optimization \CamOne{of} data partitioning strategies \CamOne{with} compute loop partitioning \CamOne{schemes}. \CamOne{\SysName applies} PIM-specific code optimizations, and \CamOne{leverages} a fast and accurate performance prediction model to select \CamOne{the best-performing code schedule for a given kernel on a target PIM architecture}.
Our evaluations in various individual ML kernels show that \SysName achieves up to \kernelmaxhbmpimgpuspeedup× speedup (\kernelavghbmpimgpuspeedup× average) on HBM-PIM, and up to \kernelmaxattaccgpuspeedup× speedup (\kernelavgattaccgpuspeedup× average) on AttAcc PIM, over GPU-only execution. In end-to-end LLM inference, \SysName on AttAcc accelerates GPT-3 and LLaMA-2 by \inferenceAvgattaccgpuspeedup× average (up to \inferenceMaxattaccgpuspeedup× \CamOne{in LLaMA-2}) over GPU. 
\revS{\SysName is open-sourced at \url{https://github.com/SPIN-Research-Group/DCC}}.
\end{abstract}
\section{Introduction}
Machine Learning (ML) models \CamOne{provide} state-of-the-art results across numerous domains including finance~\cite{heaton2017deep,ozbayoglu2020deep}, retail~\cite{oosthuizen2021artificial}, healthcare~\cite{ravi2016deep,miotto2018deep}, education~\cite{Baker2014, zhang2025machine}, entertainment~\cite{ota2017deep, zheng2024predicting}, 
and autonomous systems~\cite{tang2022perception,cordts2016cityscapes}. ML models process increasingly large datasets and comprise both compute-intensive kernels such as General Matrix-Matrix Multiplication (GEMM) and convolutions, as well as memory-intensive kernels such as General Matrix-Vector Multiplication (GEMV) and element-wise operations. 
For instance, Large Language Models (LLMs)~\cite{zhong2024distserve,su2025seesaw, Kim2025oaken, Kamath2025podattention,Li2024Foreseer,Jiang2024NEOSG}
contain  both compute-intensive fully-connected layers with GEMM kernels and memory-intensive attention layers with GEMV kernels. 
However, when executed on processor-centric CPU \CamOne{or} GPU systems, memory-intensive ML kernels are significantly bottlenecked by data movement between off-chip memory and processors~\cite{gomez2023evaluating,Gomez2022Benchmarking,giannoula2024pygim,pimdl_asplos2024,Boroumand2021Google,Boroumand2018Google}. Such memory bottlenecks increasingly limit end-to-end ML execution performance.

Processing-In-Memory (PIM)~\cite{upmem,Gomez2022Benchmarking,lee2021hardware,lee20221ynm,He2020Newton,park2024attacc,kim2025cost,li2025blockpim,liu2025mcpal,chen2025attenpim,he2025papi,li2024specpim,pimgpt_natcomm2024,lee2025paise,kim2025pimba,quinn2025longsight,pim_benchmarking_2022,pim_benchmarking_cut_2021,niu2022184qps}
has emerged as a promising paradigm to alleviate data movement bottlenecks by placing low-power processing units (\textbf{PIM cores}) near memory arrays. Numerous works~\cite{park2024attacc,kim2025cost,li2025blockpim,liu2025mcpal,chen2025attenpim,he2025papi,li2024specpim,pimgpt_natcomm2024,lee2025paise,kim2025pimba,quinn2025longsight,gomez2023evaluating,Gomez2022Benchmarking,giannoula2024pygim,pimdl_asplos2024,atim_isca2025,Boroumand2018Google,Boroumand2021Google,Yufeng2024PIM,shin2018mcdram,alves2015saving,lockerman2020livia,gao2017tetris,ke2021near,mutlu2019processing}
\CamOne{show} that PIM can provide significant performance benefits for memory-intensive ML kernels by reducing data movement costs.

A PIM system includes multiple PIM-enabled memory devices connected to a high-performance \emph{Host} processor (e.g., CPU, GPU, TPU).
Near-bank PIM devices tightly couple a PIM core with one or few DRAM banks, exploiting bank-level parallelism to enable large aggregate memory bandwidth.
Each PIM core can only access data from its local bank(s).
PIM cores of a device may not be able to directly communicate with each other, and inter-core communication can happen via Host.
Manufacturers have already started to commercialize near-bank PIM devices (referred to as \textbf{PIM backend}). UPMEM PIM~\cite{upmem,gomez2023evaluating,Gomez2022Benchmarking,pim_benchmarking_2022,pim_benchmarking_cut_2021,sparsep_pomacs2022,giannoula2024pygim} is the first commercialized near-bank PIM device, and can be integrated with CPUs. Samsung HBM-PIM~\cite{lee2021hardware} and SK Hynix GDDR6-AiM~\cite{lee20221ynm,He2020Newton}
PIM devices have been prototyped and validated, and can be integrated with GPUs.
These PIM systems can enable heterogeneous ML execution, where memory-intensive kernels run on PIM cores and compute-intensive kernels run on Host processor.

Host  and PIM cores require fundamentally different data layouts to exploit large available memory bandwidth. 
\CamOne{The} Host \CamOne{system} distributes consecutive elements \emph{across} multiple DRAM banks to exploit bank-level parallelism, and enables large bandwidth when accessing data at cache line granularity. 
In contrast, a PIM core can access data only from its local bank(s), \CamOne{and} thus in PIM, consecutive elements must be placed \emph{within} the same bank to enable efficient multi-element accesses and maximize local bandwidth.
Consequently, the PIM execution of a kernel has three steps: (1) input data rearrangements to place consecutive elements within the same banks for local PIM processing, (2) computation on PIM cores, and (3) output data rearrangements to merge partial results produced from step (2) on PIM cores or prepare output data for Host access by redistributing consecutive elements across banks. These data rearrangements are typically performed via the Host memory bus (outside PIM devices),  and thus incur large data movement costs that can dominate end-to-end performance~\cite{giannoula2024pygim,pimdl_asplos2024,Gomez2022Benchmarking,gomez2023evaluating,rhyner2024analysis}.

Therefore, programming PIM devices is a challenging task~\cite{giannoula2024pygim,pimdl_asplos2024,Gomez2022Benchmarking,gomez2023evaluating,simplepim_pact2023}. 
Programmers must manually craft data rearrangement strategies that balance data movement costs with computation efficiency, which requires deep understanding of both the PIM system and the kernel-specific access patterns of each ML kernel. \CamOne{Doing so} also demands expertise in low-level programming across multiple PIM backends \CamOne{(e.g.,} ~\cite{upmem,lee2021hardware,lee20221ynm,He2020Newton,park2024attacc,kim2025cost,li2025blockpim,liu2025mcpal,chen2025attenpim,he2025papi,li2024specpim,pimgpt_natcomm2024,lee2025paise,kim2025pimba,quinn2025longsight,gomez2023evaluating,Gomez2022Benchmarking,giannoula2024pygim,pimdl_asplos2024,atim_isca2025,gu2025pim}),
which may expose different programming interfaces and optimization capabilities.
This complexity necessitates compilation tools that automatically and intelligently generate and \CamOne{optimize \textbf{data rearrangement strategy} (i.e., how tensor data is partitioned across PIM cores) and \textbf{loop-level compute schedule} (i.e., how the computation of nested nested loops is split and assigned to PIM cores)}  to enhance programmability and minimize end-to-end execution time.


Compilation support with performance optimizations for PIM systems remains in early stages.
Existing compilation works for PIM~\cite{khan2024cinm,simplepim_pact2023,pimdl_asplos2024,pimflow_cgo2023,atim_isca2025,Xie2025UniNDP,pimcare_ics2025} lack systematic optimization and auto-tuning for diverse ML kernels and/or support for multiple PIM backends. They  largely ignore data rearrangement costs or target only UPMEM PIM.
However, UPMEM PIM is designed for CPU systems using DDR4 interfaces, has limited hardware multiplication support and lacks  floating-point arithmetic support, making it unsuitable for ML workloads that typically require GPU-PIM co-execution~\cite{park2024attacc,he2025papi,li2024specpim,kim2025pimba,giannoula2024pygim} and native floating-point operations.
ATiM~\cite{atim_isca2025} is a search-based tensor compiler that optimizes diverse ML kernels, but supports only UPMEM PIM, \CamOne{and as such is} limited to CPU-PIM co-execution.  
\cg{More critically, ATiM enables compute-centric tuning: it optimizes \CamOne{loop-level compute schedule} \emph{without} accounting for data rearrangement costs \emph{during} the \CamOne{compute schedule optimization} step.
However, as we demonstrate in §\ref{sec:motivation-datacentric},  compute code transformations and data rearrangements are \emph{interdependent}. Optimizing them in isolation yields suboptimal performance (See Fig.~\ref{fig:tvm-performance}): a compute  transformation that appears efficient in isolation may require expensive data rearrangement, while a less efficient compute transformation discarded during compute optimization could enable cheaper data rearrangements.
Achieving \CamOne{the best} performance requires balancing both costs during tuning rather than optimizing them in isolation.}

We propose \SysName, a data-centric ML compiler for PIM systems that \CamOne{incorporates} data rearrangement \CamOne{strategies into} the core of the tuning process, jointly co-optimizing them with \CamOne{loop-level compute schedules} to improve end-to-end ML kernel performance and enhance programmability. \revE{\SysName targets memory-intensive ML kernels  suitable for PIM, and has four key components.}
First, we propose a generic multi-layer abstraction that maps PIM memory hierarchy into a compute hierarchy, where PIM cores form PIM groups. This abstraction decouples the compiler from backend-specific semantics, enabling \SysName to support multiple PIM backends and explore diverse data distribution strategies.
Second, we design a data-centric schedule generator that constructs all candidate data tensor partitions of an ML kernel across PIM resources and maps them to compute loop partitions. This mapping enables comprehensive co-optimization of \CamOne{data rearrangement strategies with loop-level compute schedules}.
Third, we integrate a PIM-specific code optimizer that applies data rearrangement and compute code optimizations tailored for PIM systems. This optimizer can be extended with additional PIM-specific optimizations.
Fourth, we design a learning-based coupled \revC{performance} predictor that jointly evaluates data rearrangement and compute code execution times, and selects the best-performing end-to-end kernel configuration. This predictor provides fast and accurate performance estimates on diverse PIM backends and ML kernels.

We evaluate \SysName across diverse \revE{memory-intensive} ML kernels, tensor sizes, and models using two state-of-the-art PIM backends, HBM-PIM~\cite{lee2021hardware} and AttAcc~\cite{park2024attacc}. In individual ML kernels, \SysName \CamOne{provides} up to \kernelmaxhbmpimgpuspeedup× speedup (\kernelavghbmpimgpuspeedup× average) on HBM-PIM and up to \kernelmaxattaccgpuspeedup× speedup (\kernelavgattaccgpuspeedup× average) on AttAcc compared to GPU-only execution. In end-to-end LLM inference on AttAcc, \SysName \CamOne{provides} up to \inferenceMaxattaccgpuspeedup× speedup (\inferenceAvgattaccgpuspeedup× average) over GPU and up to \inferenceMaxattaccBaselinespeedup× speedup (\inferenceAvgattaccBaselinespeedup× average) over AttAcc's original implementation.


Overall, we make the following contributions: 
\begin{itemize}[topsep=0pt,leftmargin=8pt,nosep,partopsep=0pt]
\item We demonstrate that data rearrangement \CamOne{strategies} and \CamOne{loop-level compute schedules} must be jointly optimized for \revE{memory-intensive} ML kernels on PIM systems, and propose \SysName, the first data-centric ML compiler that co-optimizes both in a unified tuning process.
\item We design a multi-layer abstraction that maps PIM memory hierarchy into a compute hierarchy, and propose a schedule generator that constructs data tensor partitions and maps them to compute loop partitions. We integrate PIM-aware optimizations for data rearrangements and compute code, and employ a learning-based coupled \revC{performance}  predictor to select best-performing end-to-end  \CamOne{code schedule for an ML kernel on a target PIM backend}.
\item We evaluate \SysName on two state-of-the-art PIM backends, and show that \SysName \CamOne{provides} significant performance improvements \CamOne{over best prior works} for diverse ML kernels, tensor sizes, and LLMs.
\item \revS{We open-source \SysName in our GitHub repository: \url{https://github.com/SPIN-Research-Group/DCC}}.
\end{itemize}

\section{Background and Motivation}\label{sec:background-motivation}

\subsection{Processing-In-Memory (PIM) Architectures}\label{sec:background-pim}

Processing-In-Memory (PIM)~\cite{Gomez2022Benchmarking,lee2021hardware,lee20221ynm,He2020Newton, mutlu2020modern, mutlu2019enabling, upmem, Mutlu2025Memory, mutlu2019processing, he2025papi, park2024attacc}
places low-power processing units  near memory arrays, and can alleviate data movement bottlenecks in processor-centric CPU/GPU systems. Near-bank PIM designs tightly couple each core with one (or a few) DRAM banks that can \CamOne{supply data to the core}. Near-bank PIM provides larger aggregate memory bandwidth and parallelism compared to near-rank PIM\cite{mutlu2020modern, ke2021near}, where cores are placed at DRAM buffer chip. UPMEM PIM~\cite{upmem,Gomez2022Benchmarking} is the first commercialized PIM system. Samsung HBM-PIM~\cite{lee2021hardware}, Alibaba HB-PNM~\cite{niu2022184qps} and SK Hynix GDDR6-AiM~\cite{lee20221ynm,He2020Newton} have also been prototyped. 

UPMEM PIM is built on DDR4 interfaces for CPU systems, has limited hardware multiplication support and no floating-point arithmetic units~\cite{Gomez2022Benchmarking}. Due to these limitations, UPMEM PIM with CPU-PIM co-execution in ML workloads cannot typically outperform  GPU-only execution~\cite{Gomez2022Benchmarking,giannoula2024pygim,pimdl_asplos2024,sparsep_pomacs2022,rhyner2024analysis}.
In contrast, Samsung HBM-PIM~\cite{lee2021hardware}, Alibaba HB-PNM~\cite{niu2022184qps} and SK Hynix GDDR6-AiM~\cite{lee20221ynm,He2020Newton} are 3D PIM memory devices that \CamOne{can} integrate with a high-performance \textbf{xPU} processor such as \CamOne{a} GPU \CamOne{or} TPU, provide hardware multiplication and floating-point units (e.g., FP16), \CamOne{which makes} them suitable for ML acceleration. Building on these industry products, numerous research works~\cite{park2024attacc,kim2025cost,li2025blockpim,liu2025mcpal,chen2025attenpim,he2025papi,li2024specpim,pimgpt_natcomm2024,lee2025paise,kim2025pimba,quinn2025longsight}  explore enhanced PIM core microarchitectures \CamOne{and system designs} to further accelerate ML kernels. In this work, we target near-bank PIM devices that can be integrated with GPUs/TPUs, have hardware multiplication support  and/or floating-point arithmetic units.

\begin{figure}[h]  
    \centering 
    \includegraphics[width=0.9\linewidth]{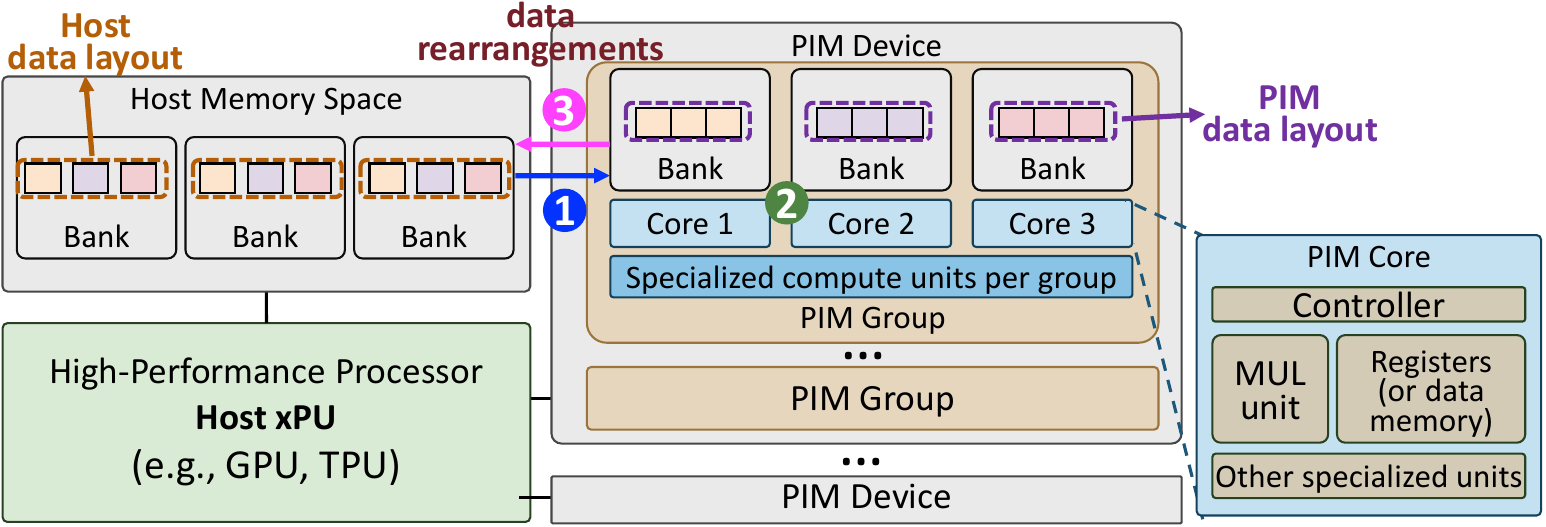}
    \caption{Near-bank PIM architecture and kernel execution workflow: (1) input rearrangement, (2) computation execution, and (3) output rearrangement steps.}
    \label{fig:pim-arch-overview}
\end{figure}

We find common characteristics in existing  PIM designs~\cite{upmem,lee2021hardware,lee20221ynm,He2020Newton,park2024attacc,kim2025cost,li2025blockpim,liu2025mcpal,chen2025attenpim,he2025papi,li2024specpim,pimgpt_natcomm2024,lee2025paise,kim2025pimba,quinn2025longsight,pim_benchmarking_2022,pim_benchmarking_cut_2021,niu2022184qps} 
as shown in Fig.~\ref{fig:pim-arch-overview}.
First, a high-performance processor, \textbf{Host xPU} (e.g., GPU, TPU) with on-chip cache hierarchy and Host memory typically connects to multiple PIM devices.
Second, each PIM device contains multiple processing units (\textbf{PIM cores}) that can be organized into \textbf{PIM groups}. Each PIM core (e.g., a SIMD unit, GEMV unit) 
has exclusive access to one or few local memory banks, enabling larger aggregate memory bandwidth and lower latency than Host cores have.
Third, PIM cores have register files or scratchpad data memory (referred to as \textbf{data memory}). They may support SIMD execution, hardware multiplication, low-precision floating-point or other specialized operations.
Fourth, the Host sends PIM instructions that are stored in register files (or instruction cache) of PIM cores. PIM cores execute PIM instructions to move data from/to their local banks to/from registers and/or perform computations (e.g., MUL, ADD) using data stored in registers.
Fifth, PIM designs may integrate specialized compute units per PIM group for specific ML kernels. For example, AttAcc~\cite{park2024attacc} provides hardware softmax and accumulator support per PIM group, enabling full attention  execution on the PIM side. 
Finally, PIM cores may not be able to directly communicate with each other, and communication between them typically happens via the Host memory bus.

PIM devices \revE{typically operate in two modes. Either Host xPU executes a  kernel (e.g., compute-intensive kernel) and uses PIM devices as standard DRAM memory for loads/stores. Or,  Host xPU offloads a  kernel (e.g., memory-intensive kernel) to PIM cores, and PIM cores execute the kernel by accessing their local DRAM banks exclusively. We find that} Host xPU and PIM cores require different data layouts to fully leverage their available memory bandwidth, \CamOne{which necessitates} software-managed data rearrangements (Fig.~\ref{fig:pim-arch-overview}). Host xPU distributes consecutive elements \emph{across} multiple DRAM banks to exploit bank-level parallelism for large bandwidth, when accessing multiple elements as a  cache line. 
Instead, PIM cores require consecutive elements \emph{within} the same bank: since each core accesses data only from its local bank(s), maximizing local bandwidth requires placing consecutive elements in the same bank to efficiently fetch multiple elements at once as a block.
Thus, PIM kernel execution has three steps (Fig.~\ref{fig:pim-arch-overview}): \textcircled{1}  input data rearrangements to place consecutive elements within the same bank for local PIM core processing, \textcircled{2} computation execution on PIM cores, and \textcircled{3} output data rearrangements to either merge partial results from step \textcircled{2} on PIM cores or prepare output data for Host access by redistributing consecutive elements across multiple banks.
\CamOne{Data rearrangements} are typically performed via the Host memory bus (outside PIM devices), thus incurring significant data movement overheads.

\subsection{Need for Data-Centric ML Compiler Support for PIM Architectures}\label{sec:motivation-datacentric}
PIM software support remains in early stages, with limited automation and compilation frameworks.
SimplePIM~\cite{simplepim_pact2023} optimizes only 1D tensor kernels on UPMEM PIM. PIM-DL~\cite{pimdl_asplos2024} effectively supports only the GEMM kernel. CINM~\cite{khan2024cinm} integrates multi-level intermediate representations to lower abstractions to PIM, again targeting UPMEM. PIMFlow~\cite{pimflow_cgo2023} supports only convolution kernels, automating kernel offloading to GPU or PIM, but without kernel-level tuning optimizations and largely ignoring data rearrangement costs.
These works lack systematic optimization and auto-tuning for diverse ML kernels and support for multiple PIM backends. Moreover, most of them target UPMEM PIM, a CPU-integrated DDR4-based device that has 
no floating-point support \CamOne{and weak multiplication support}. This makes UPMEM unsuitable for ML models, where compute-intensive kernels typically run on GPUs/TPUs and memory-intensive kernels on PIM cores may need floating-point arithmetic and \CamOne{strong multiplication units}.

ATiM~\cite{atim_isca2025} is a search-based tensor compiler designed for UPMEM PIM. 
ATiM tunes ML kernels on CPU-PIM systems, however, CPU-PIM co-execution \CamOne{on UPMEM systems} typically performs worse than GPU-only execution~\cite{giannoula2024pygim,Gomez2022Benchmarking, pim_benchmarking_2022,pim_benchmarking_cut_2021,sparsep_pomacs2022}. In contrast, GPU-PIM co-execution can deliver substantial performance benefits in ML~\cite{park2024attacc,he2025papi,li2024specpim,kim2025pimba} over GPU-only \CamOne{execution}.
Although ATiM considers data rearrangement costs, it could not  \CamOne{achieve best} performance due to its \emph{compute-centric} tuning approach. ATiM has a three-step sequential process: \CamOne{it} (i) uses TVM~\cite{Chen2018TVM} to find and fix a set of templates for compute \CamOne{schedule} generation, (ii) generates data rearrangements needed for each template using UPMEM-specific optimizations, and (iii) searches within this \emph{fixed} set of templates to find the best-performing  configuration.
This sequential approach is suboptimal, because it treats compute \CamOne{schedule optimization}  and data rearrangement as \emph{independent} problems, while they are fundamentally \emph{interdependent}. A compute schedule that appears efficient in isolation may require expensive data rearrangements, while a slightly less efficient schedule might enable cheaper data rearrangement costs, yielding better end-to-end performance. By fixing compute templates before generating data rearrangements, ATiM explores only a restricted search space and cannot discover configurations where alternative compute schedules paired with different data transformations could minimize total time.

\begin{figure}[!htb]
    \centering
    \includegraphics[width=\linewidth]{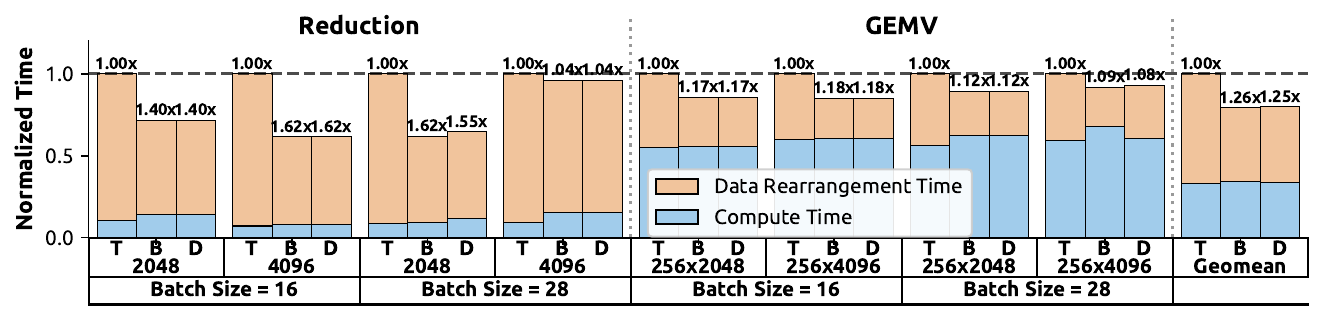}    
    \caption{Normalized breakdown of compute and data rearrangement times in Reduction and GEMV kernels, comparing the TVM-based compilation scheme (\textbf{T}), a manually-tuned best-performing implementation (\textbf{B}) and \revE{\SysName (\textbf{D})} using various matrix sizes. The numbers on each bar show speedup over \textbf{T}.}
    \label{fig:tvm-performance}
\end{figure}

Fig.~\ref{fig:tvm-performance} shows the breakdown of compute (step (2) in Fig.~\ref{fig:pim-arch-overview}) and data rearrangement (steps (1) and (3) in Fig.~\ref{fig:pim-arch-overview}) times for the Reduction and GEMV kernels running on AttAcc PIM~\cite{park2024attacc} (See §\ref{sec:methodology}) using various matrix sizes. We evaluate \revE{a TVM-based~\cite{Chen2018TVM} compilation scheme (\textbf{T}) that fairly mirrors ATiM's approach~\cite{atim_isca2025}, against a manually-tuned best-performing implementation (\textbf{B}) and \CamOne{our proposal}  \SysName (\textbf{D}).}
For the TVM-based scheme, we tune TVM via performance profiling of Reduction and GEMV on the PIM devices. TVM selects the best-performing compute \CamOne{schedule} templates based on its cost model. For each template, we generate the optimized data rearrangements and measure end-to-end performance (compute time plus data rearrangement time). 
We present 
the configuration with the best performance among TVM's selected templates.
\cg{For the manually-tuned best-performing implementation, we manually select some promising data partitioning strategies across PIM cores, and deploy an optimized compute \CamOne{schedule} for all of them. Then, we evaluate all of them and present the configuration that \CamOne{provides} the minimum total execution time.}


\newcommand{\tvmBestAllAvgbaselineSpeedup}{\revC{1.28}}
\newcommand{\tvmBestcompAvgbaselineSlowdown}{\revC{1.16}}
\newcommand{\tvmBestdataAvgbaselineSpeedup}{\revC{1.55}}
\newcommand{\tvmDCCAllAvgbaselineSpeedup}{1.27}

We make two observations. First, the TVM-based approach has suboptimal performance. \CamOne{It is} on average \tvmBestAllAvgbaselineSpeedup$\times$ worse  than the \CamOne{best manual} implementation. This is \revE{because the TVM-based approach (followed by ATiM~\cite{atim_isca2025}) is compute-centric}: it selects compute \CamOne{schedule} templates that minimize compute time, while largely ignoring data rearrangement costs. Consequently, data rearrangements \CamOne{are responsible for} $64.68\%$ of the total kernel time on average.
Second, while the manually-tuned implementation has \tvmBestcompAvgbaselineSlowdown$\times$ worse compute time than  the TVM-based approach, it reduces data rearrangement costs by \tvmBestdataAvgbaselineSpeedup$\times$ compared to the TVM-based scheme. Thus, it provides better trade-offs that result in better end-to-end performance. Overall, these results indicate that compute schedules and data rearrangements are \emph{interdependent}: \CamOne{obtaining the best} performance requires balancing both costs. 
\revE{\SysName jointly co-optimizes data rearrangement costs and compute schedules, and \CamOne{provides} \tvmDCCAllAvgbaselineSpeedup$\times$ over the compute-centric TVM-based approach}, \CamOne{performing very similarly to the best manual implementation}.

\section{\SysName: Overview}
\begin{figure}[t]  
    \centering 
    \includegraphics[width=0.99\linewidth]{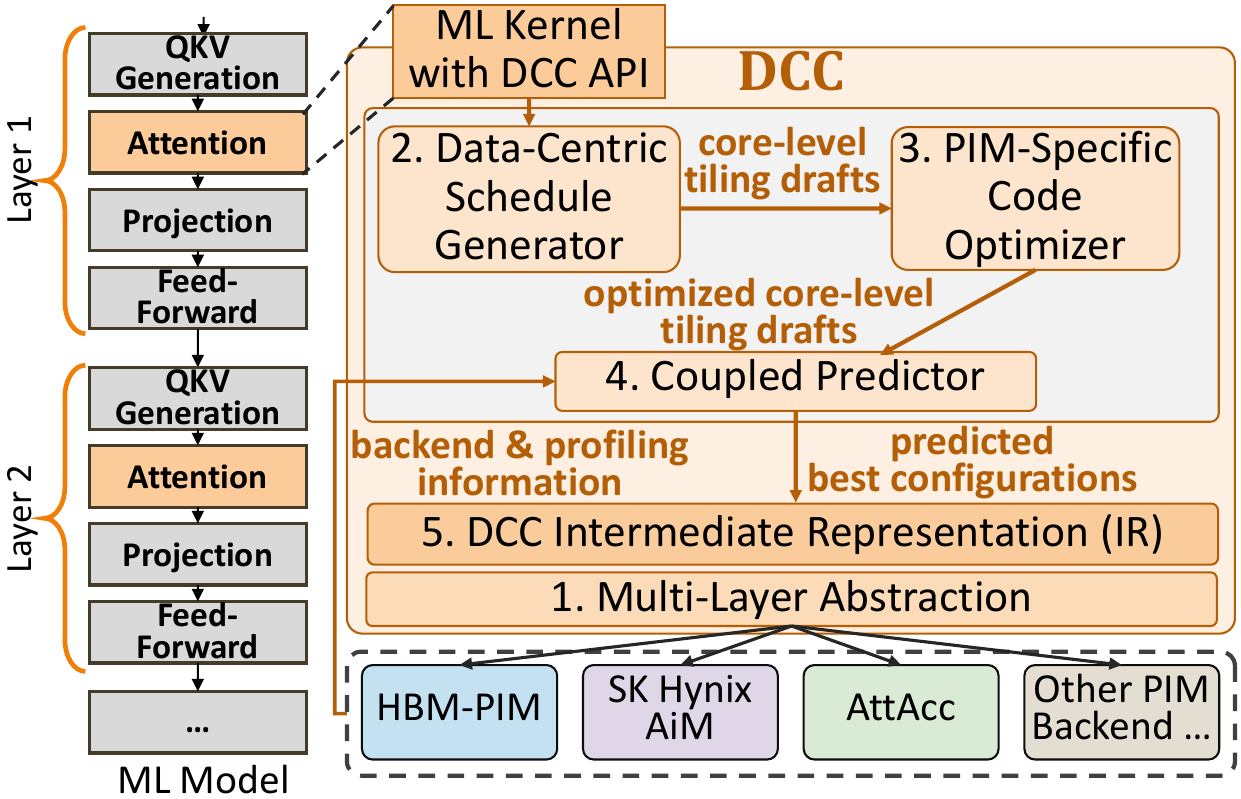}
    \caption{\ArXiv{High-level Overview of \SysName.} }
    \label{fig:pim-dcc-overview-abstraction}
\end{figure}

\SysName (Fig.~\ref{fig:pim-dcc-overview-abstraction}) is the first data-centric ML compiler for PIM \CamOne{systems} that \emph{co-optimizes} data rearrangement strategies \CamOne{together} with compute \CamOne{schedule} in a unified tuning process. \SysName supports diverse ML kernels and multiple PIM backends. Fig.~\ref{fig:pim-dcc-overview-abstraction} presents a high-level overview of \SysName components. Users develop ML kernels using the \SysName API \CamOne{(§\ref{sec:mechanism-interface})}, which enables execution on target PIM backends. At compile time, \SysName analyzes ML kernels in the model and trains its coupled predictor. At runtime, once input tensor dimensions are known (e.g., token sizes in LLMs), \SysName uses its pre-trained  predictor to \CamOne{select optimized configurations} for all PIM-running kernels. \CamOne{The end-to-end kernel configuration} orchestrates data loading to the target PIM backend, computation on PIM cores, and final output data with the appropriate layout. \SysName has \ArXiv{five} components.
 

\noindent\textbf{1. Multi-Layer Abstraction.} 
We design a general multi-layer abstraction for near-bank PIM systems, where PIM cores form PIM groups, \CamOne{which} enables \SysName to reason about data distribution across memory banks and data processing on PIM groups and cores. Our abstraction (i) enables coarse-grained optimizations at the PIM group level and fine-grained optimizations at the PIM core level, and (ii) decouples the compiler from backend-specific semantics, allowing \SysName to support multiple current and future PIM backends. \newline
\noindent\textbf{2. Data-Centric Schedule Generator.} We follow a \emph{data-first} scheduling strategy that first generates all candidate \CamOne{data partitions} across PIM resources, and then maps them to corresponding loop partitions in the \CamOne{compute schedule}, creating \emph{tiling drafts}. This approach enables effective \emph{co-optimization} of data rearrangement with \CamOne{loop-level compute schedule}. \newline
\noindent\textbf{3. PIM-Specific Code Optimizer.} \CamOne{We introduce PIM-specific performance optimizations for data rearrangement strategy and loop-level compute schedule of each tiling draft.}
This optimizer can be easily extended with additional backend-agnostic optimizations or backend-specific passes to create specialized compiler variants for particular PIM backends. \newline
\noindent\textbf{4. Coupled \revC{Performance} Predictor.} We use a learning-based \revC{performance} predictor that models the interdependence between data rearrangement costs and compute efficiency and selects the best-performing configuration that minimizes end-to-end \revC{execution}  time. It can provide fast and accurate performance predictions on various PIM backends. \newline
\noindent\textbf{5. Intermediate Representation.} \ArXiv{We design a data-centric Intermediate Representation (IR) that captures the optimized data partitioning strategy and loop-level compute schedule derived from the best-performing configuration. Given an ML kernel written in high-level Python as input, the IR encodes how data is distributed across PIM cores and how computation is structured within them. This intermediate layer is subsequently lowered to backend-specific PIM instructions, enabling support for multiple PIM backends.}
\newline

\section{\SysName: Detailed Design}

\subsection{Multi-Layer Abstraction for PIM Architectures}
\vspace{-5pt}

To efficiently support multiple PIM backends, we introduce a \textbf{general multi-layer abstraction}, which maps the traditional memory hierarchy into a compute hierarchy, allowing \SysName to configure where and how each ML tensor is distributed and processed. 
\revE{Our abstractions covers various near-bank PIM backends~\cite{upmem,Gomez2022Benchmarking,lee2021hardware,lee20221ynm,He2020Newton,park2024attacc,kim2025cost,li2025blockpim,liu2025mcpal,chen2025attenpim,he2025papi,li2024specpim,pimgpt_natcomm2024,lee2025paise,kim2025pimba,quinn2025longsight,pim_benchmarking_2022,pim_benchmarking_cut_2021,niu2022184qps}, 
that have the common characteristics described in §\ref{sec:background-pim}.
These PIM backends} have similar logical organization: they place one PIM core (e.g., specialized acceleration or floating point unit) close to one or a few memory banks. Multiple PIM cores can form \textbf{PIM groups} (e.g., cores of the same memory channel or the same DRAM rank can form a group), and the memory bus controller of Host xPU can serve as a controller for all PIM groups.
PIM devices may include specialized compute units for each PIM core group (Fig.~\ref{fig:pim-arch-overview}) that can access data from 
multiple memory banks within the same PIM group.
Although the PIM core may have different compute units across different PIM backends, 
the overall computational model remains consistent across PIM backends. Key differences lie primarily in hardware circuitry design and core microarchitecture rather than execution semantics.
Any current or future PIM system that satisfies our proposed multi-layer abstraction can be seamlessly supported by \SysName for ML kernel acceleration. Our abstraction has three levels. \CamOne{Two of these levels are within the} PIM device.

\begin{figure}[h]  
   \centering 
   \includegraphics[width=0.99\linewidth]{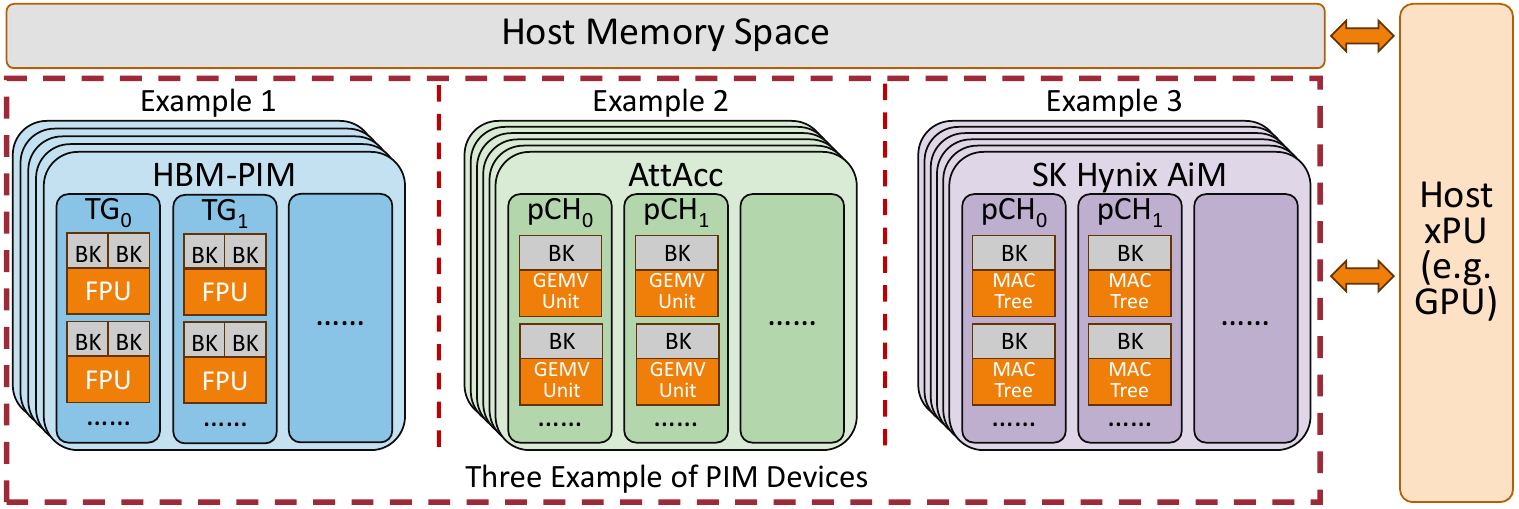}
   \caption{\revD{Example PIM backends instantiated under \SysName's multi-layer abstraction. }}
   \label{fig:abstraction-example}
\end{figure}

\noindent\textbf{(1) System Level.}
Our system-level PIM abstraction  consists of three major components (See Fig.~\ref{fig:pim-arch-overview}): the \textbf{Host xPU}, \textbf{Host memory space} and the \textbf{PIM devices}. The Host xPU (e.g., CPU, GPU, TPU) orchestrates global execution control, data rearrangements, and kernel scheduling. PIM devices perform near-memory computation. The Host memory space is an address space that follows the Host-side data layout, accessible through standard DRAM commands (e.g., LD/ST). It maps to physical addresses in PIM-enabled or normal memory banks.
\revA{For example (Fig.~\ref{fig:abstraction-example}), in the HBM-PIM system~\cite{lee2021hardware}, the system level includes the GPU device as Host xPU and multiple HBM-PIM devices. In the AttAcc PIM system~\cite{park2024attacc}, it includes the GPU as Host xPU and  multiple AttAcc PIM devices.}


\noindent\textbf{(2) PIM Group Level.}
A PIM group represents multiple PIM cores (and their local memory banks) belonging to the same memory channel or rank, depending on the device architecture. A PIM group may include specialized compute units that access data from all banks within the group and/or buffers to temporarily store group-wide data.
\revA{For instance, in HBM-PIM~\cite{lee2021hardware}, the group level could include multiple FPU units of one or multiple Thread Groups (TGs) (Fig.~\ref{fig:abstraction-example}) of the same HBM-PIM device. In AttAcc PIM~\cite{park2024attacc},  it could include multiple GEMV and accumulator units of one or multiple pseudo-Channels (pCHs) of the same AttAcc PIM device.}
Each group is managed by the Host xPU memory controller, which may issue group-level instructions. A group-level instruction can be a read/write/compute command (i) that broadcasts to all cores within the group, typically using identical bank address offsets, or (ii) is executed by the specialized compute units of the group. In the latter case, the instruction may access data from shared group-wide buffers or merge partial results produced by PIM cores of the group.
Each group has a dedicated memory bus to the Host xPU, enabling \SysName to schedule \emph{parallel} data rearrangements across multiple PIM groups. 
Host can coordinate global data movements across \CamOne{PIM} groups or between Host and PIM memory. 
By scheduling coarse-grained group-level instructions broadcast to all cores within a \CamOne{PIM} group, \SysName can reduce instruction traffic on the memory controller, and enable parallel operations \CamOne{in the PIM} group's cores.

\noindent\textbf{(3) PIM Core Level.}
A PIM core represents the processing unit and its local memory bank(s). 
It can include a SIMD unit or  Floating-Point Unit (FPU) or any specialized compute unit. It may have   scratchpad memory or register files to temporarily store data for processing.  
\revA{For example, in HBM-PIM~\cite{lee2021hardware}, the core level includes one single core consisting of an FPU and its register files placed close two (neighboring) memory banks. In AttAcc PIM~\cite{park2024attacc}, it includes one GEMV unit placed close to its memory bank.}
If the Host memory space is mapped to PIM banks, a  portion of each bank serves as Host memory, and the PIM backend provides address offsets for the PIM mapping.
PIM cores are controlled via  bank-level instructions or broadcast group-level instructions sent by the Host memory controller, or instructions stored in dedicated instruction memory per core. 
By scheduling fine-grained bank-level instructions, 
\SysName enables optimizations on individual \CamOne{PIM} cores.

This abstraction enables PIM devices to operate under either group-level or bank-level control for near-bank computation. 
It also enables \SysName to optimize at multiple granularities: coarse-grained at the PIM group level (e.g., effectively partitioning data across groups or scheduling computation on specialized units of a group) and fine-grained at the PIM core level (e.g., optimizing memory access within a bank or local data layout and computation).
It decouples \SysName  from hardware-specific instruction semantics, \CamOne{thereby} enabling \SysName to easily support multiple PIM backends that conform to this abstraction. 


\subsection{Data-Centric Schedule Generation}\label{sec:mechanism-generation}
Traditional ML compilers~\cite{Chen2018TVM,feng2023tensorir,Ding2023Hidet,xing2022bolt,Gupta2025SPLAT,Liu2025CROSS,Ahrens2025Finch,Du2025SRSparse,won2023unified,Ye2023SparseTIR,Kjolstad2017TACO} use loop transformations to divide computation into blocks, and improve data locality. 
\SysName adopts a \textbf{data-first schedule} strategy that inverts this process: it \CamOne{1)} generates all candidate \revD{data partitions of 
data tensors (called \textbf{data tiles}) across PIM groups and cores (following its multi-layer abstraction), \CamOne{2)} maps data tiles to their corresponding loop partitions (ranges of loop variables) in the compute code (called \textbf{compute tiles}), \CamOne{3)} optimizes each \CamOne{configuration}'s performance, \CamOne{4)} then uses its predictor to evaluate data-compute tile mappings, \CamOne{5)} selects the best-performing schedule, and \CamOne{6)} generates PIM instructions for the best-performing \CamOne{end-to-end kernel configuration} for the target PIM backend. 
}

\begin{figure}[ht]  
    \centering 
    \includegraphics[width=\linewidth]{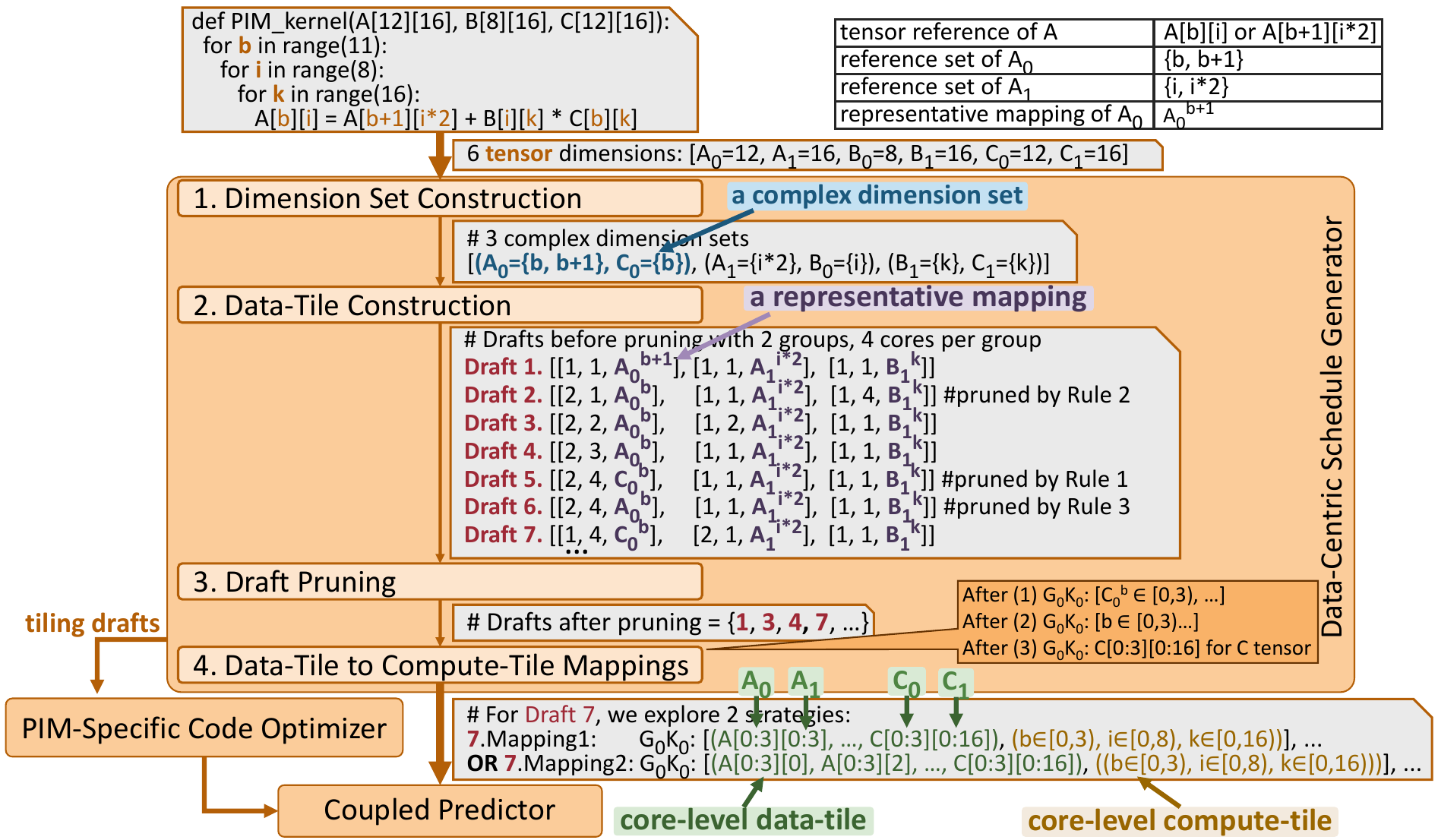}
    \caption{Example  schedule generation process for an ML kernel.}
    \label{fig:pim-dcc-example}
\end{figure}

To find the \CamOne{best} data-compute schedule for a given kernel and hardware configuration, \SysName generates a comprehensive set of \textbf{tiling drafts}, each representing a potential mapping between tensor sizes, computation loops, and PIM hardware resources. The generation process includes four stages shown in Fig.~\ref{fig:pim-dcc-example} for an example ML kernel and explained next.

\noindent\textbf{1. Dimension Set Construction.}
To jointly co-optimize compute-data, we need to correlate the data tensors with the \CamOne{loop-level compute schedule}.
For a $k$-dimensional tensor $A$, we map its dimensions $A_0$, $A_1$, ... $A_k$ to the \CamOne{loop-level compute schedule}: we leverage loop variables as intermediaries to associate tensor dimensions to \CamOne{mapping functions}. 
Specifically, for a tensor $A$ with tensor reference $A[b$+$1][i$*$2]$, where $A_0$ is the first dimension and $A_1$ is the second dimension, we use the variables and constants to build mapping functions for each dimension: e.g., the mapping functions $F(b)$=$b$+$1$ and $F(i)$=$i$*$2$ are for \CamOne{the} dimensions $A_0$ and $A_1$, respectively. A tensor may have multiple tensor references in  \CamOne{nested loops}, e.g., tensor $A$ in Fig.~\ref{fig:pim-dcc-example} has two references $A[b][i]$ and $A[b$+$1][i$*$2]$, creating multiple mapping functions for the same tensor dimension. The dimension $A_0$ is associated with both $F(b)$=$b$ and $F(b)$=$b$+$1$. We build all mapping functions for each tensor dimension and collect them into a \textbf{reference set}. The reference set for dimension $A_0$ is $\{b, b$+$1\}$, and for $A_1$ is $\{i, i$*$2\}$. The reference set for the tensor dimension $C_0$ is $\{b\}$, and for $C_1$ is $\{k\}$. \CamOne{Next,} when two or more tensor dimensions share the \emph{same} loop variable in their reference sets, they are grouped into a \textbf{complex dimension set}. For example, dimensions $A_0$ and $C_0$ both contain variable $b$ in their reference sets, \CamOne{thus} we group them as $(A_0$=$\{b,b+1\}, C_0$=$\{b\})$ to be processed jointly, since these dimensions must share \CamOne{the same loop indexing values, i.e., $b$ takes the same indexing values in loop iterations.} 
Tensor dimensions that do not share variables with any other dimension form simpler dimension sets.

\noindent\textbf{2. Data-Tile Construction.}
We explore all candidate data tile options through exhaustive depth-first search with memoization, \CamOne{constructing} data tiles for each dimension set. For the $i$-th dimension set among $n$ dimension sets, given $k$ available PIM groups and $m$ available cores per group, we try all possible PIM allocations, i.e., $[1,k]$ PIM groups and $[1,m]$ cores per group. The available PIM groups and cores for the ($i$+$1$)-th dimension set are computed recursively as $avail\_resources_{i+1}$=\scalebox{0.8}{$\frac{avail\_resources_i}{allocated\_resources_i}$}.
Memoization caches all valid allocation solutions for each subproblem, 
enabling reuse when the same PIM resource configuration \CamOne{is encountered again} during search.
Once the allocation of PIM resources is done for all dimension sets, we create multiple tiling drafts as follows. Each draft has $n$ parts, one for each dimension set. Each part includes three parameters: [the number of groups, the number of cores, a representative mapping]. The \textbf{representative mapping} is an assignment of a mapping function (e.g., $b$+$1$) to a tensor dimension (e.g., $A_0$). We create drafts via exhaustive search 
for all possible assignments of representative mappings for a given dimension set.
For instance, in Fig.~\ref{fig:pim-dcc-example}, the first draft has three parts for its three dimension sets. The first part  is [1, 1, $A_0^{b+1}$], which represents 1 PIM group, 1 core per group, and the representative mapping $A_0^{b+1}$. The second part [1, 1, $A_1^{i*2}$] represents 1 group, 1 core per group, and the representative mapping $A_1^{i*2}$. Similarly, the third part is [1, 1, $B_1^{k}$]. 
When this step finishes, we have generated all possible tiling drafts for all possible data tiles.

\noindent\textbf{3. Draft Pruning.}\label{sec:mechanism-pruning}
\SysName applies \CamOne{three} pruning rules to eliminate unpromising tiling drafts and reduce tuning time, while maintaining high performance.
\textbf{Rule 1.} We remove redundant tiling drafts that have the same data tiles and equivalent representative mappings for all dimension sets. In Fig.~\ref{fig:pim-dcc-example}, the 5th draft [[2, 4, $C_0^b$]...] and the 6th draft [[2, 4, $A_0^b$]...] are redundant. In their first part (\CamOne{which} relates to the first dimension set), they have the same data \CamOne{partition}, i.e., 2 PIM groups and 4 cores per group, and equivalent representative mappings, i.e., their representative mappings $C_0^b$ and $A_0^b$ use the same mapping function $F(b)$=$b$, \CamOne{and} thus they \CamOne{generate \emph{identical} data partitions}.  
The second and third parts of these tiling drafts are identical.
Thus, one of these two drafts, i.e., the 5th draft, is removed.
\textbf{Rule 2.} PIM backends~\cite{lee2021hardware, lee20221ynm, He2020Newton, chen2025attenpim,li2024specpim, park2024attacc, pimgpt_natcomm2024,he2025papi}
support SIMD-style instructions that enable PIM cores to fully leverage the large memory bandwidth available in their local bank(s). For example, HBM-PIM~\cite{lee2021hardware} supports 16-way SIMD instructions. In a PIM backend with $d$-way \CamOne{SIMD} instruction support, the draft pruner speculatively eliminates drafts where the \CamOne{size} of the dimension set assigned to a PIM core is not a multiple of $d$: these drafts cannot fully utilize the core's local bandwidth and hardware datapaths designed for $d$-way execution.
In Fig.~\ref{fig:pim-dcc-example}, let us assume a PIM backend with 16-way MUL instructions, and consider the dimension set ($B_1$=\{$k$\}, $C_1$=\{$k$\}) for the computation $B[i][k] \times C[b][k]$. The 2nd draft \CamOne{assigns} 4 cores to the dimensions $B_1$ and $C_1$ \CamOne{(the first dimensions of tensors $B$ and $C$), both} of size 16, \CamOne{so each core must process 4 elements along the dimension assigned to it.} Since 4 is not a multiple of the 16 \CamOne{(16-way MUL instruction)}, this draft is pruned. If all drafts fail to satisfy the $d$-way \CamOne{instruction} alignment requirement, or if the PIM backend does not support \CamOne{SIMD} instructions, \SysName skips this pruning rule.
\textbf{Rule 3.} \CamOne{Within a PIM group, execution performance is determined by the worst-performing core in the group. When comparing multiple drafts, if the worst per-core performance is the \emph{same} across all corresponding PIM groups for each part of the draft, those drafts are execution-performance equivalent. In this case, only one draft needs to be kept, while the remaining drafts are pruned. To apply this pruning rule, we estimate per-PIM-core performance using the dimension partition size assigned to each core. This partition size is derived from the dimension included in the representative mapping of each part of the draft.
The PIM core with the largest dimension partition size within a PIM group is taken as the worst-case performance representative for that group.
In Fig.~\ref{fig:pim-dcc-example}, the 4th and 6th drafts each have three parts. For the first part, whose representative mapping is $A_0^{b}$, both drafts have 2 PIM groups. The \CamOne{worst per-PIM-core performance for} the  $A_0$ dimension derived from $A_0^{b}$  for the corresponding PIM groups in both drafts is the \emph{same}: in the 4th draft, it is round(\scalebox{0.8}{$\frac{A_0}{groups \times cores}$})=round(\scalebox{0.8}{$\frac{12}{2\times3}$})=2, and in the 6th draft, it is round(\scalebox{0.8}{$\frac{A_0}{groups \times cores}$})=round(\scalebox{0.8}{$\frac{12}{2\times4}$})=2. The remaining two parts of the 4th and 6th drafts yield the same worst per-core performance across all their PIM groups.  Since the two drafts are execution-performance equivalent, one is arbitrarily pruned, i.e.,  the 6th draft is pruned.} 


\noindent\textbf{4. Data-Tile to Compute-Tile Mappings.}
Once the drafts are pruned, \SysName constructs core-level \textbf{data-tiles} and maps them to core-level \textbf{compute-tiles}. 
The \CamOne{remaining} drafts include only the high-level tiling \CamOne{of each part, i.e., the number of PIM groups, PIM cores per group, and the representative mapping of the corresponding dimension set}. However, we need to (1) lower the high-level tiling to core-level data-tiles, (2) map the core-level data-tiles to core-level compute-tiles, \revS{and (3) determine the necessary data indices of all tensors corresponding to loop variable values of the compute-tiles.}

\noindent\textbf{(1)} \CamOne{For each part of a draft (which corresponds to a dimension set), we lower the high-level tiling to core-level data-tiles. Given the number of PIM groups, PIM cores per group, and the representative mapping of a dimension set, the data-tile for core $K_j$ of the PIM group $G_i$ is computed by dividing the size of the dimension (the dimension is derived from the representative mapping)  by the product of the number of PIM groups and the number of PIM cores per group, i.e., \scalebox{0.8}{$\frac{dimension\_size}{groups \times cores}$}. In Fig.~\ref{fig:pim-dcc-example}, the first part of the 7th draft [1, 4, $C_0^b$] has 1 PIM group and 4 PIM cores per group, and the representative mapping $C_0^{b}$ (which corresponds to the dimension set ($A_0$=\{$b$,$b$+$1$\}, $C_0$=\{$b$\})). The data-tile size for each core along the dimension $C_0$ is \scalebox{0.8}{$\frac{C_0}{groups \times cores}$}=\scalebox{0.8}{$\frac{12}{1\times4}$}=$3$.  Thus, the data-tile for the PIM core $K_0$ of the PIM group $G_0$ for the first dimension set is $C_0$ $\in$ $[0$,$3)$ (Fig.~\ref{fig:pim-dcc-example}). We similarly compute the data-tiles for all parts of the draft (corresponding to all dimension sets) and for all cores.}


\noindent\textbf{(2)} \CamOne{For each part of the draft (which corresponds to a dimension set), we map the core-level data-tile to a core-level compute-tile by determining the loop range of the associated variable for the kernel loops. Given the data-tile of core $K_j$ of PIM group $G_i$ and the mapping function (e.g., $F(b)$=$b$) derived from the representative mapping, the compute-tile is the set of all values of the associated variable (e.g., $b$) for which the mapping function result falls within the data-tile (e.g., $F(b)$ $\in [0$,$3)$). In Fig.~\ref{fig:pim-dcc-example}, for the first part of the 7th draft, core $K_0$ of the PIM group $G_0$ has data-tile $[0$,$3)$ (from step (1)) for the first dimension set, and the mapping function is $F(b)$=$b$.  The compute-tile is thus the set of values of $b$ satisfying $F(b)$ $\in [0$,$3)$, which gives $b$ $\in$ $[0$,$3)$. 
Similarly, we calculate all loop ranges for all loop variables across all cores. } 


\noindent\textbf{(3)} Once the compute-tile for each core is determined, \SysName \revS{derives the \textbf{tensor data indices} \CamOne{(e.g., A[$0$:$3$][$0$:$3$])}} \CamOne{that each core must access for \textbf{all} the tensor dimensions, corresponding to the loop variable values of the compute-tile}.
\CamOne{In Fig.~\ref{fig:pim-dcc-example}, for the 7th draft, \SysName derives tensor data indices for all tensor dimensions $A_0$, $A_1$, $B_0$, $B_1$, $C_0$ and $C_1$}.
When \CamOne{the required} indices of a tensor are non-contiguous along a dimension, \SysName \CamOne{explores} two alternative strategies.
\CamOne{Strategy} (i) expands the indices of this dimension to cover a contiguous range, adding the missing or straddling values \CamOne{among the required indices} (e.g., in Fig.~\ref{fig:pim-dcc-example} in \textit{7.Mapping1} the indices of $A_1$ are expanded from \{$0$, $2$\} to $[0$:$3]$). \CamOne{Strategy} (ii) keeps only the \CamOne{required} indices  \CamOne{as-is} (e.g., in \textit{7.Mapping2} the index set of $A_1$ is [$0$] and [$2$] (kept as it is)). 
\CamOne{Since each strategy has different performance implications, \SysName generates both and selects the better-performing one.}

\subsection{PIM-Specific Code Optimizer}
Once tiling drafts are generated, data \CamOne{is} mapped across PIM banks and compute loop partitions  \CamOne{is} mapped to PIM cores. \SysName then applies a \textbf{PIM-specific code optimizer} for \emph{each} tiling draft that further optimizes \revB{performance. We exploit common PIM architectural characteristics to enable  one optimization that reduces data rearrangement costs and one that improves the \CamOne{execution efficiency of loop-level compute schedule}.} 

\noindent\textbf{I) Data Rearrangement Optimization.}
Host and PIM \CamOne{cores} require different data \CamOne{layouts} to achieve high performance. In Host memory, contiguous data needs to be distributed across multiple memory channels to exploit channel-level parallelism. For example, on a two-channel CPU, a 1KB block is divided into 16 cache \CamOne{blocks} (indexed 0-15), with odd-indexed lines \CamOne{mapped} to one channel and even-indexed lines to the other channel. This enables \emph{parallel} sequential reads across multiple memory channels.
\CamOne{In contrast}, PIM \CamOne{cores} require contiguous data to be stored in a single memory bank, i.e., using one single memory channel for writes, since each PIM core can only process data stored in its local bank(s). This creates the following data \CamOne{rearrangement} challenge: (i) reading contiguous data sequentially from Host memory exploits Host channel parallelism, but \CamOne{results in} writes \CamOne{that use} only one PIM memory channel, underutilizing PIM bandwidth, while instead (ii) using multiple PIM channels for parallel writes requires simultaneous reads from \emph{non-contiguous} Host addresses, potentially causing Host channel conflicts and degrading read performance.
To enable  channel parallelism for both Host reads and PIM writes, \SysName exploits the controllable on-chip memory (e.g., GPU shared memory) to reorganize data within the Host. For a PIM device with $N$ channels, the compiler calculates the data block size as $B = \frac{\text{on-chip memory size}}{N}$. \SysName then serially reads $N$ blocks of size $B$ from Host memory to on-chip memory, followed by parallel write operations that distribute these blocks across all $N$ PIM channels simultaneously. This approach transforms  data movement into $N$ sequential read operations from  Host channels 
and $N$ parallel writes to PIM channels, 
leveraging channel parallelism in both reads and writes.
If there is no controllable on-chip memory in the system (e.g., on CPUs), \SysName performs data rearrangement using the PIM backend's default data copy or DMA interfaces. For PIM backends with specialized layout constrains (e.g., alignment or interleaving requirements), \SysName \CamOne{applies} additional \CamOne{compliation passes for perfmoring required data layout transformations}.

\noindent\textbf{II) Compute Code Optimization}. PIM backends~\cite{lee2021hardware,lee20221ynm,He2020Newton,park2024attacc} can use specialized DRAM commands to control computation \revB{on PIM cores. However, when all PIM cores perform computations} in parallel, the Host memory controller must issue significantly more commands than in conventional DRAM, \revB{creating a control bottleneck due to the limited memory bus bandwidth.} 
To address this, PIM backends~\cite{lee2021hardware, lee20221ynm,He2020Newton,park2024attacc} introduce group-level commands that provide SIMD-style control: \revB{instead of issuing individual compute commands to each PIM core, a single compute command is broadcast to \emph{all} cores within a \emph{group}, with each core operating on its own local data, significantly reducing command bus overhead.} \SysName employs a hierarchical command generation strategy leveraging both \textbf{group-level} and \textbf{bank-level} control to balance efficiency and flexibility.
\SysName \revB{detects and prioritizes \textbf{group-level commands} wherever possible.} 
Since DRAM command formats constrain group-level commands to use identical address offsets for all cores in the group, \SysName pads each core's tensor data to the same size \revB{to ensure} consistent local addressing.
This way \SysName allows the memory controller to issue one command per PIM group, significantly reducing command traffic.
If the backend lacks group-level command support or when cores need to access different address offsets, 
\SysName generates \emph{parallel} \textbf{bank-level commands}, 
achieving fine-grained, bank-level parallelism.  This hierarchical command generation strategy adapts to different PIM backends, and minimizes command traffic.

Our optimizer applies PIM-specific optimizations that accelerate performance while remaining backend-agnostic. It has a modular design that facilitates the integration of additional optimizations shared across multiple PIM backends. Developers can also extend \SysName with custom optimization passes tailored to specific PIM backends, creating specialized compiler variants (e.g., \SysName+HBM for HBM-PIM-specific optimizations).

\subsection{Coupled \revC{Performance}  Predictor}
Although pruning substantially reduces \CamOne{the number candidate drafts}, the schedule generator can still produce thousands of valid tiling drafts. Profiling all drafts on \CamOne{the target} PIM system can be prohibitively expensive. To efficiently identify the \CamOne{best} draft, \SysName employs a \textbf{coupled learning-based \revC{performance} predictor} that estimates end-to-end execution time.
The coupled predictor serves two purposes: (i) co-estimating data rearrangement and compute costs to find the \CamOne{best} tiling draft, and (ii) providing fast and accurate predictions across multiple PIM backends. 
\revE{We choose a learning-based model (already effectively adopted  by widely-used ML compilers~\cite{Chen2018TVM, zheng2020ansor,chen2018learning}) that can be easily trained (retrained or fine-tuned) for a new PIM backend, and provide both \CamOne{low latency} and high prediction accuracy. We use an XGBoost model~\cite{chen2016xgboost} to predict end-to-end kernel time, because XGBoost has been proven highly effective for compiler cost modeling in prior works~\cite{Chen2018TVM, zheng2020ansor, chen2018learning, shao2022tensor, ganesan2020case, ahn2020chameleon, mishra2022compoff}.

While analytical models can achieve sufficient accuracy with PIM device-specific formulas and parameters, they are difficult to generalize across multiple PIM backends with competitive accuracy and maintain as PIM architectures evolve.
Moreover, unlike neural predictors (e.g., \CamOne{these} used in ~\cite{zheng2020ansor}), XGBoost is lightweight in memory and compute requirements.} 


The coupled predictor supports both static and dynamic tensor sizes. For static sizes, 
where model and input dimensions are fixed during inference, \SysName trains the predictor offline and configures the best-performing schedule during model initialization (similarly to prior ML compilers~\cite{Chen2018TVM, zheng2020ansor, chen2018learning}). For dynamic inputs, which occur in LLMs, \SysName finds and selects the best-performing schedules at inference time.

\noindent\textbf{\emph{Offline Training:}}
Given model-defined or user-provided tensor sizes, \SysName samples a diverse subset of drafts across different tensor sizes from the schedule generation step. Each draft 
is profiled on the target PIM backend to measure execution time as labels, with draft's configurations and backend information as inputs to the XGBoost model. After training, \SysName predicts performance for all drafts at given tensor sizes of ML kernels and records the best-performing draft in a lookup table. When multiple drafts have identical predicted performance, one is selected randomly.
This training occurs \emph{once} per PIM backend using given ML models and tensor sizes. 
Offline training for all ML kernels across all tensor sizes, ML models, and PIM backends used in our evaluation takes only \kerneltrainingtimeshort seconds.
This cost is negligible and is amortized across multiple users' inference requests for that ML model.
At runtime, \SysName \emph{directly} uses the best-performing draft for recorded tensor sizes.

\noindent\textbf{\emph{Dynamic Prediction:}}
When a new tensor size appears during inference, \SysName generates the corresponding tiling drafts and uses the predictor to estimate performance, recording the best-performing schedule in the lookup table for future use. 
In ML models with dynamic tensor sizes, e.g., evaluation of LLMs in Fig.~\ref{fig:inference}, the cost to generate drafts on-the-fly and estimate performance is accounted for in total time. 
Optionally, \SysName can support training the predictor at inference time: it can profile a small batch of new drafts, and use the (updated) predictor to estimate performance for remaining drafts.

\subsection{Programming Interface \& Intermediate Representation (IR)}\label{sec:mechanism-interface}

\SysName provides a Python~\cite{python2016} interface with PyTorch~\cite{paszke2019pytorch}, described in Table~\ref{tab:dcc_interface}, to deploy, integrate and compile \CamOne{ML kernels for PIM execution.}
\revS{\SysName compiles Python kernels marked with the \textit{@\SysName\_kernel} annotator. The programmer writes a Python kernel that describes tensor computations, specifying tensor shapes and data types needed at compile time.  All parameters and return values must be PyTorch tensors or scalar constants. The kernel body describes computations over tensors using affine loop nests, elementary arithmetic operators, and backend-supported instructions, including operations on tensors of varying numeric precision (e.g., integer and floating-point) and conditional operations. If there are multiple loop nests that share the same mapping function and loop range, \SysName merges them using the same complex dimension set.} 

{
\setlength{\tabcolsep}{2pt}
\begin{table}[h]
\centering
\resizebox{0.999\linewidth}{!}{
\begin{tabular}{|l||l|}

\hline
\rowcolor{teal!16}  \textbf{Interface} & \textbf{Description} \\
\hline
\hline
\textit{@\SysName\_kernel} & \makecell[l]{Annotator to define a PIM-running kernels.} \\ 
\hline
\textit{\SysName.Tensor} & \makecell[l]{Tensor class used in kernel.It can be  conve-\\rted to \textit{torch.Tensor}.}\\
\hline
\textit{\SysName.Layer} & \makecell[l]{Class to define a new PIM-running layer.} \\
\hline
\textit{\SysName.Kernel} & \makecell[l]{Class to represent a PIM-running kernel.} \\
\hline
\textit{\SysName.initKernel()} & \makecell[l]{Function to initialize the kernel for given\\input sizes. It returns a \texttt{\SysName.Kernel}.} \\
\hline
\textit{\SysName.Kernel.update()} & \makecell[l]{Function called when the best tiling draft \\ has been determined.} \\
\hline
\textit{\SysName.Kernel.preLoad()} & \makecell[l]{Function to load data to PIM devices and\\return a PIM address. It also supports part-\\ial updates of tensors in PIM memory with \\ address offset.}\\
\hline
\textit{\SysName.Kernel()} & \makecell[l]{Function to execute a PIM-running kernel\\ with \textit{torch.Tensor} or PIM address.} \\
\hline
\textit{\SysName.setModel()} & \makecell[l]{Function to register the model and extract\\ model-level metadata necessary for tuning.} \\
\hline
\end{tabular}
}
\caption{The \SysName \revS{Python} Programming Interface.}
\label{tab:dcc_interface}

\end{table}
}

Users can integrate \CamOne{a kernel executed on PIM into an ML model} using \textit{\SysName.Layer} to create ML layers that replace existing model layers. PIM-running kernels must be initialized with
input tensor sizes. \SysName provides \textit{\SysName.Tensor} and \textit{\SysName.Tensor.zero()} to create and initialize temporary tensors. 
After adding \CamOne{PIM-accelerated} layers to the model, \SysName initializes necessary model metadata, infers all tensor sizes, and performs offline training.
At inference, when a request is received, \SysName asynchronously infers tensor sizes for all kernels executed on PIM, and selects their best-performing schedules. Users can override the function \textit{\SysName.Layer.update()} to pre-load data to PIM devices using \textit{\SysName.Kernel.preLoad()} before running the kernel. When \textit{forward()} is called, the kernel is executed on PIM cores with the input data and returns the results as a \textit{torch.Tensor}. 
\SysName synergistically works with xPU compilers (e.g., ML compilers~\cite{Chen2018TVM,Ding2023Hidet} for GPUs) via PyTorch. 
During the optimization for xPU, the PIM layer will be configured 
as non-fusible to enable intra- and inter-kernel optimizations. 
\SysName  compiles and optimizes this layer for the PIM backend.

\begin{figure}[h]  
    \centering 
    \includegraphics[width=0.999\linewidth]{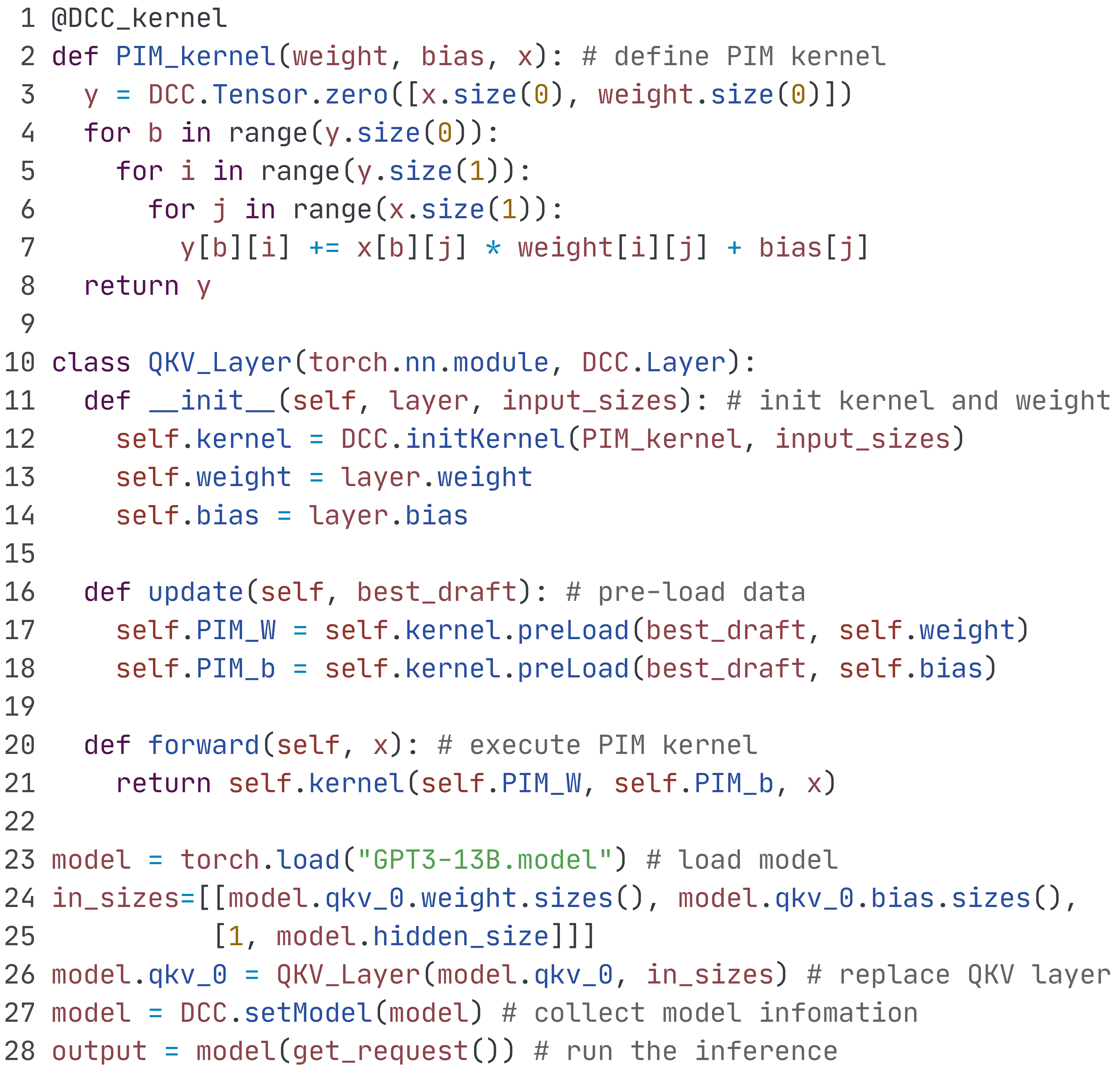}
    \caption{An example of adding \SysName kernels to a GPT3 model.}
    \label{fig:pim-dcc-code}
\end{figure}

Fig.~\ref{fig:pim-dcc-code} shows an example of replacing the QKV generation layer in GPT-3 13B model with a PIM-accelerated layer. Lines 1-8 define the kernel \CamOne{(running on PIM)} with for loops. Line 10 creates class \textit{QKV\_Layer} inheriting from the \texttt{torch.nn.Module} and \texttt{\SysName.Layer}. In lines 11-14, \SysName initializes the kernel and model parameters. Lines 16-18 pre-load weights and bias to PIM devices after having selected the best-performing schedule. Line 21 executes the kernel with the given input and pre-loaded data. Lines 23-28 replace the original QKV layer in the loaded GPT-3 model and handle inference requests.

\SysName \revS{parses Python kernels marked  \textit{@\SysName\_kernel},  compiles them to its \emph{data-centric} Intermediate Representation (IR), and then generates \CamOne{low-level backend instructions} based on the hardware intrinsics supported by the target PIM backend. If the target PIM backend does not support a specific operation, \SysName reports a compilation error. The \SysName's IR constructs are described in Table~\ref{tab:dcc_ir-constructs}. } 

\setlength{\tabcolsep}{2pt}
\begin{table}[h]
\vspace{1pt}
\centering
\resizebox{0.999\linewidth}{!}{
\begin{tabular}{|l||l|}

\hline
\rowcolor{teal!16}  \textbf{IR Construct} & \textbf{Description} \\

\hline
\hline
{@hardware\_platform[G,C]} &  \makecell[l]{Defines the PIM hardware configurati-\\on for the annotated DCC kernel func-\\tion, i.e., the number of PIM groups \\and PIM cores per PIM group.} \\
\hline
{\makecell[l]{func @name(\%arg:$<$shape$\times$\\dtype$>$): 
$<$shape$\times$dtype$>$}} & \makecell[l]{Defines a PIM kernel function with ty-\\ped input/output.} \\
\hline
{\makecell[l]{complex\_dimset \%ds \\ \{\%dim[0],...\}}} &  \makecell[l]{Constructs a complex dimension set as\\ \%ds with one or more mapping functi-\\ons \%dim[idx].}\\
\hline
{\makecell[l]{\%dim[i] = \%t.dim[axis]\\ \textrightarrow expr}} & \makecell[l]{Declares a mapping function \%dim[i]\\  on a tensor dimension \%t.dim[axis] as\\  $expr$.} \\
\hline
{[representative]} & \makecell[l]{Marks a mapping function in a compl-\\ex dimension set as the representative\\  mapping.} \\
\hline
{\%ds.range = (start, end, step)} & \makecell[l]{Specifies the iteration range of the lo-\\op variable in a complex dimension set \\\%ds.} \\
\hline
data\_tile \%dt: (g,c)\textrightarrow  (g,c,ts) &  \makecell[l]{Constructs a data tile \%dt as the numb-\\er of PIM groups($g$), PIMcores ($c$), and \\ the generated tile size ($ts$). $g$ and $c$\\ are included in the draft description.} \\
\hline
{\makecell[l]{compute\_tile \%ct: (g,c,ts)\\ \textrightarrow [[$R_0$,...],...]}} &  \makecell[l]{Generates a compute tile \%ct based on \\ the data tile \%dt. Each \%ct is a 2D arr-\\ay containing the loop variable ranges \\$R_i$ of all $i$ partitions of a tensor dime-\\nsion.} \\
\hline
{\makecell[l]{tensor\_tile \%tt: \%ct\\ \textrightarrow[[$T_0$,...], ...]}} & \makecell[l]{Generates all necessary data indices of\\ all tensors from a compute tile. Each \\\%tt is a  2D array describing the necess-\\ary indices  $T_i$ of all  partitions of a ten-\\sor dimension.}\\
\hline
{\%t\_core = rearrange \%t\textrightarrow \%tt} & 
\makecell[l]{Rearranges tensor data between Host\\ and PIM data layout in local banks.\\ \%t\_core contains the core-level local\\ data reference of the tensor \%t.} \\
\hline
{[to\_PIM\_core/to\_host]} & 
\makecell[l]{Shows the direction of  data movement. } \\
\hline
{\makecell[l]{parallel\_PIM (groups, cores) \\ \{...\}}} & \makecell[l]{Launches parallel execution across all\\ PIM groups and PIM cores concurren-\\tly.} \\
\hline
{\makecell[l]{loop ($x_1$,...,$x_n$) in (\%$ct_1$,...,\\\%$ct_n$) \{...\}}} &  \makecell[l]{Iterates over the compute-tiles \%$ct_1$,...,\\\%$ct_n$ using $n$ loop variables $x_1$,...,$x_n$ to\\ drive PIM core-local computation.} \\
\hline
\end{tabular}
}
\caption{IR constructs of \SysName  and their descriptions.}
\label{tab:dcc_ir-constructs}
\end{table}

\begin{figure}[h]  
    \centering 
    \includegraphics[width=\linewidth]{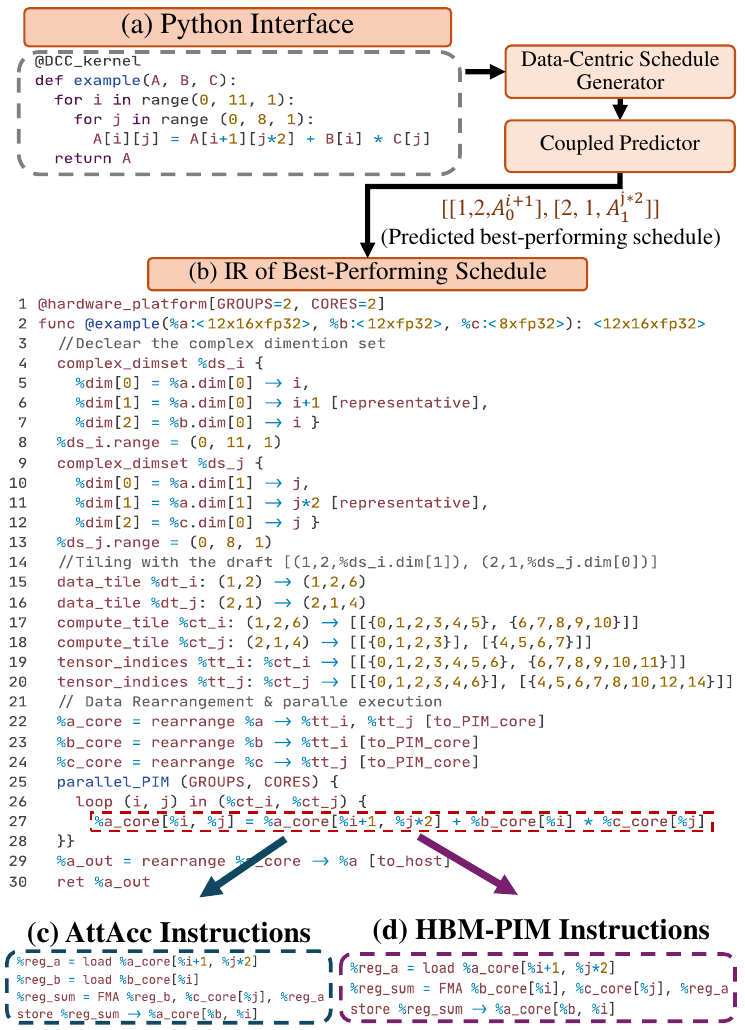}
    \caption{Compilation flow of a kernel using \SysName: from the front-end code to its IR of data- and compute-tiles, and then to backend instructions.}
    \label{fig:pim-ir-code}
\end{figure}

Fig.~\ref{fig:pim-ir-code} \revS{illustrates how \SysName compiles a kernel (Fig.~\ref{fig:pim-ir-code}a) from the \CamOne{Python} code to its IR of data- and compute-tiles for a given schedule, and subsequently to final backend instructions. For a given schedule produced by the schedule generator (§\ref{sec:mechanism-generation}), the IR generator produces an IR program (Fig.~\ref{fig:pim-ir-code}b). Lines 1--2 declare the available PIM hardware resources and the typed input and output tensors. Lines 4--13 create the \emph{complex dimension sets} with designated \emph{representative mappings} (§\ref{sec:mechanism-generation}.1), and declare their iteration ranges (lines 8, 13). Lines 15--16 generate the \emph{data-tiles} by binding the dimension sets to PIM hardware resources (groups, cores) (§\ref{sec:mechanism-generation}.2) and calculating the per-core data-tile sizes (§\ref{sec:mechanism-generation}.4 step (1)).
Lines 17--18 derive the \emph{compute-tiles} from the \emph{data-tiles}, and generate the iteration range for each loop variable at the core level (§\ref{sec:mechanism-generation}.4 step (2)).
Lines 19--20 derive the necessary \emph{tensor data indices} at the core level for all tensors, corresponding to loop variable values (§\ref{sec:mechanism-generation}.4 step (3)), and implementing the~(ii) tensor indexing strategy in this example. Lines 22--29 derive the end-to-end execution flow (See Fig.~\ref{fig:pim-arch-overview}): performing input data rearrangements for the input tensors (lines 22--24), executing the kernel computation concurrently (parallel) across all PIM cores in tiled compute domains (lines 25--28), and performing output data rearrangements for the output tensor results (line 29).
Once IR generation is complete, \SysName parses the IR program and translates it to the target backend instructions: Fig.~\ref{fig:pim-ir-code}c and Fig.~\ref{fig:pim-ir-code}d show the generated backend instructions of the tensor computation of line~27 of Fig.~\ref{fig:pim-ir-code}b for the AttAcc and HBM-PIM backends, respectively.}

\section{Evaluation}
\label{sec:evaluation}

\label{sec:methodology}

\begin{figure*}

\centering

\begin{subfigure}{0.35\textwidth}
    \centering
    \includegraphics[width=\linewidth]{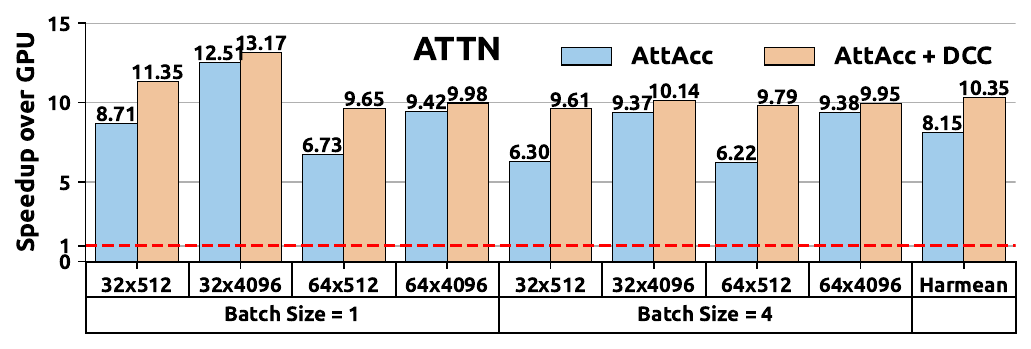}
    \caption*{}
    \label{fig:speedup_attaccattn_a}
\end{subfigure}~
\begin{subfigure}{0.35\textwidth}
    \centering
    \includegraphics[width=\linewidth]{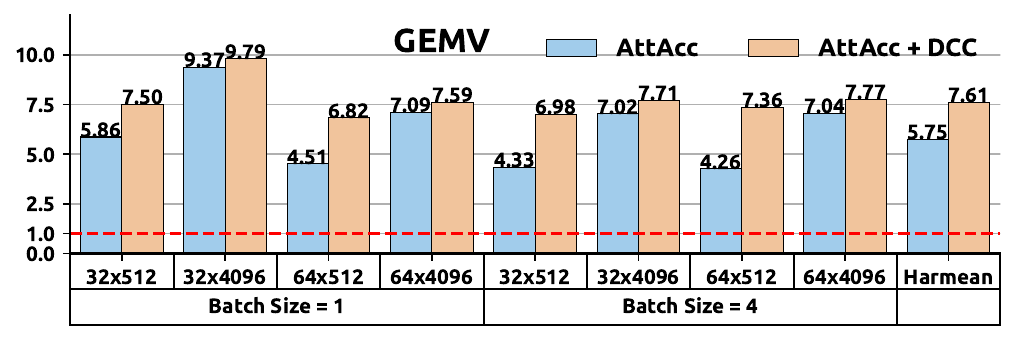}
    \caption*{}
    \label{fig:speedup_attaccattn_b}
\end{subfigure}~
\begin{subfigure}{0.3\textwidth}
    \centering
    \includegraphics[width=\linewidth]{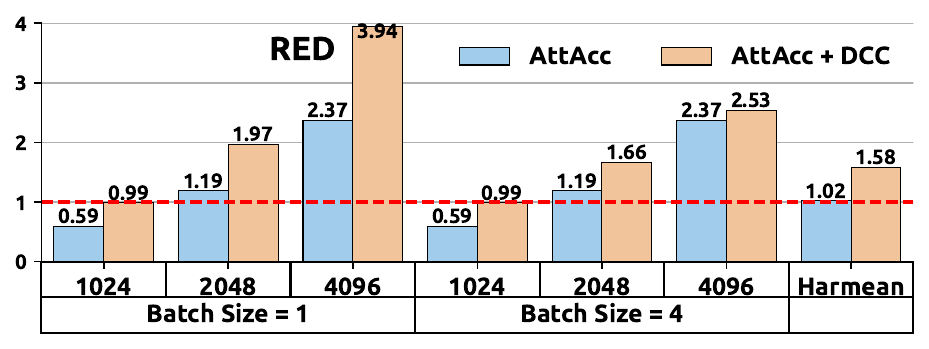}
    \caption*{}
    \label{fig:speedup_attaccattn_c}
\end{subfigure}

\vspace{-\baselineskip}

\caption{Speedup of AttAcc and AttAcc+\SysName over GPU for ATTN, GEMV and RED kernels, when varying the tensor sizes.}
\label{fig:attacc_speedups}
\end{figure*}

\noindent\textbf{Simulation Methodology.} We modify and use the open-source AttAcc simulator~\cite{park2024attacc} with Ramulator 2.0~\cite{luo2023ramulator2}. We evaluate data movement costs to/from PIM devices using DRAM commands (i.e., LD/ST) simulated in Ramulator 2.0. We modified the simulator to support \revC{two PIM backends, i.e.,} AttAcc~\cite{park2024attacc} and HBM-PIM~\cite{lee2021hardware}, with its corresponding DRAM compute commands~\cite{lee2021hardware}. 
We simulate a heterogeneous platform  with a NVIDIA A100 GPU and \revC{5 PIM-enabled HBM-type devices}, corresponding to 80GB GPU memory. \revC{Our platform configuration is  consistent with prior state-of-the-art works~\cite{park2024attacc,he2025papi}.} We evaluate HBM3 memory with 5.2Gbps per pin and running at 333MHz. \revA{Each HBM has 16 pseudo-channels \revC{(pCH)}, 2 ranks per pCH, 2 banks groups per rank and 4 banks per bank group (64 banks per pCH). We follow the timing parameters that AttAcc used: tCK=0.79, tRCD=19, tRP=19, tCL=19, tCCD=4, BL=2.} In AttAcc, each bank \revC{of each HBM-type device} is equipped with a GEMV unit and each channel has a softmax unit. In HBM-PIM, every two banks \revC{of each HBM-type device} share a 16-way FP16 FPU and two 16$\times$256-bit GRF registers (one per bank).

\noindent\textbf{ML Kernels and Models.}
We evaluate seven memory-intensive ML kernels. In AttAcc, we evaluate the general-matrix-vector-multiplication (\textbf{GEMV}), reduction (\textbf{RED}) and attention (\textbf{ATTN}) kernels. In HBM-PIM, we evaluate the GEMV, RED, vector addition (\textbf{VA}) and \textbf{RELU} kernels. \cg{Attention requires a softmax unit which only exists on AttAcc. However, AttAcc does not have near-bank compute units that could support and run RELU and VA.}  For GEMV and ATTN, \revC{we use input size 128, the most common per-head dimension in LLMs.} We also evaluate end-to-end inference using GPT3-13B and LLAMA2-33B models. We evaluate FP16 data type.

\noindent\textbf{Comparison Points.}
\label{sec:comparison}
We evaluate five comparison points. (1) \textbf{GPU}: \revB{all kernels run on an A100 GPU configured in AttAcc and Ramulator 2.0 simulators}. (2) \textbf{AttAcc}: AttAcc's~\cite{park2024attacc} default open-source implementation that distributes different batches or attention heads across 16 pCHs, partitions the first tensor dimension across 16 bank groups per pCH, and partitions the second tensor dimension across 4 banks per bank group. (3) \textbf{AttAcc+\SysName}: we enable \SysName compiler on AttAcc.
(4) \textbf{HBM-PIM}: we use AttAcc's data distribution. 
(5) \textbf{HBM-PIM+\SysName}: we enable \SysName compiler on HBM-PIM.

For offline training, \revB{we configure XBoost with learning\_rate=0.1, max\_depth=8, num\_boost\_round=5000. For all workloads, we use 15\% of the pruned drafts as training samples, and 85\% as test set.}
The \revC{compilation time (offline draft generation, training, prediction) for all evaluated workloads is \kerneltrainingtimeshort  seconds in total using a system with 32-core AMD EPYC 7513 CPU, 128GB DDR4 memory and NVIDIA A100 GPU. }

\subsection{ML Kernel Performance}
\label{sec:performance}

Fig.~\ref{fig:attacc_speedups} shows the speedup of AttAcc and AttAcc+\SysName over the GPU baseline in various ML kernels, when varying the batch size, number of heads and tensor sizes.

\newcommand{\gemvMaxAttaccGpuspeedup}{9.79}
\newcommand{\attnMaxAttaccGpuspeedup}{13.17}
\newcommand{\redMaxAttaccGpuspeedup}{3.94}
\newcommand{\gemvMaxAttaccBaselinespeedup}{1.73}
\newcommand{\attnMaxAttaccBaselinespeedup}{1.57}
\newcommand{\redMaxAttaccBaselinespeedup}{1.66}


\newcommand{\gemvAvgAttaccGpuspeedup}{\revC{7.61}}
\newcommand{\attnAvgAttaccGpuspeedup}{\revC{10.35}}
\newcommand{\redAvgAttaccGpuspeedup}{\revC{1.58}}
\newcommand{\gemvAvgAttaccBaselinespeedup}{\revC{1.26}}
\newcommand{\attnAvgAttaccBaselinespeedup}{\revC{1.23}}
\newcommand{\redAvgAttaccBaselinespeedup}{\revC{1.48}}
\newcommand{\redSmallAvgAttaccBaselinespeedup}{0.59}

We make three key observations. 
\revC{First, AttAcc significantly outperforms GPU across most kernels, achieving \attnAvgAttaccGpuspeedup$\times$ for ATTN, \gemvAvgAttaccGpuspeedup$\times$ for GEMV, and \redAvgAttaccGpuspeedup$\times$ for RED on average, with the exception of RED at tensor size 1024 where GPU performs better.}
AttAcc leverages high aggregate PIM bandwidth and integrates specialized units (GEMV, softmax, and accumulator) that provide hardware-level support for these operations.
Second, \SysName provides further performance improvements over AttAcc: \attnAvgAttaccBaselinespeedup$\times$, \gemvAvgAttaccBaselinespeedup$\times$, and \redAvgAttaccBaselinespeedup$\times$ average speedup, and up to \attnMaxAttaccBaselinespeedup$\times$, \gemvMaxAttaccBaselinespeedup$\times$, and \redMaxAttaccBaselinespeedup$\times$ peak speedup for ATTN, GEMV, and RED kernels, respectively. Notably, for RED with tensor size 1024, AttAcc underperforms GPU by \redSmallAvgAttaccBaselinespeedup$\times$, while with \SysName achieves almost same performance with GPU.
Third, with \SysName performance scales well in RED as tensor size increases.
In RED, data rearrangement costs dominate the total time (see Fig.~\ref{fig:execution_breakdown_all} in §\ref{eval:kernel-breakdown}) and with larger tensor sizes, \SysName explores a larger search space, allowing it to more effectively optimize data rearrangement costs. In RED with batch size 1 and tensor size 4096, \SysName provides a large speedup of 3.94$\times$.
Overall, \SysName significantly accelerates end-to-end time on the state-of-the-art AttAcc  backend by up to \kernelmaxattaccgpuspeedup$\times$  (\kernelavgattaccgpuspeedup$\times$ on average) over GPU across diverse ML kernels, batch and tensor sizes, thanks to its comprehensive data-compute co-optimization.

\newcommand{\gemvMaxHbmpimGpuspeedup}{7.68}
\newcommand{\redMaxHbmpimGpuspeedup}{3.35}
\newcommand{\vaMaxHbmpimGpuspeedup}{5.36}
\newcommand{\reluMaxHbmpimGpuspeedup}{6.7}

\newcommand{\gemvMaxHbmpimBaselinespeedup}{2.36}
\newcommand{\redMaxHbmpimBaselinespeedup}{2.34}
\newcommand{\vaMaxHbmpimBaselinespeedup}{1.98}
\newcommand{\reluMaxHbmpimBaselinespeedup}{1.78}



\newcommand{\gemvAvgHbmpimGpuspeedup}{\revC{5.59}}
\newcommand{\redAvgHbmpimGpuspeedup}{\revC{1.35}}
\newcommand{\vaAvgHbmpimGpuspeedup}{\revC{2.07}}
\newcommand{\reluAvgHbmpimGpuspeedup}{\revC{2.66}}
\newcommand{\gemvAvgHbmpimBaselinespeedup}{\revC{1.53}}
\newcommand{\redAvgHbmpimBaselinespeedup}{\revC{2.09}}
\newcommand{\vaAvgHbmpimBaselinespeedup}{\revC{1.64}}
\newcommand{\reluAvgHbmpimBaselinespeedup}{\revC{1.54}}
\newcommand{\redSmallAvgHbmpimGpuSlowdown}{\revC{0.48}} 
\newcommand{\redSmallAvgHbmpimBaselineSpeedup}{\revC{2.26}}

\begin{figure}[ht]

\captionsetup[subfigure]{labelformat=empty}

\begin{subfigure}{0.47\textwidth}
    \includegraphics[width=\linewidth]{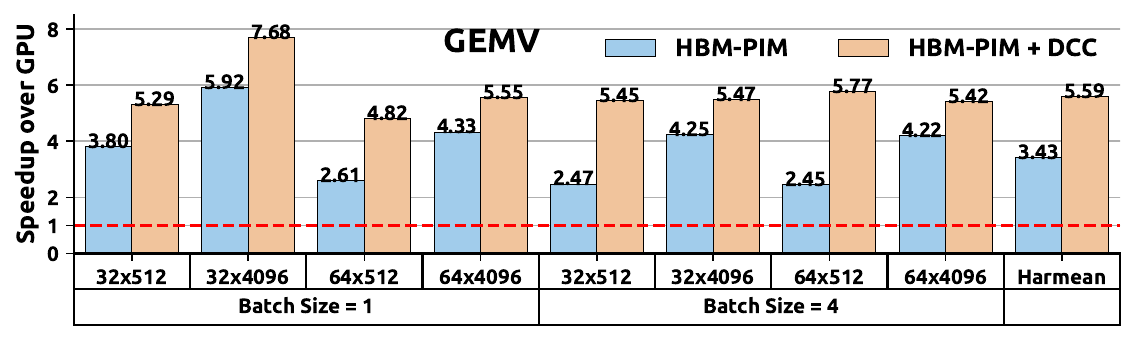}
    \label{fig:speedup_attaccattn_d}
\end{subfigure}

\vspace{-\baselineskip}
\vspace{-0.05cm}

\begin{subfigure}{0.47\textwidth}
    \includegraphics[width=\linewidth]{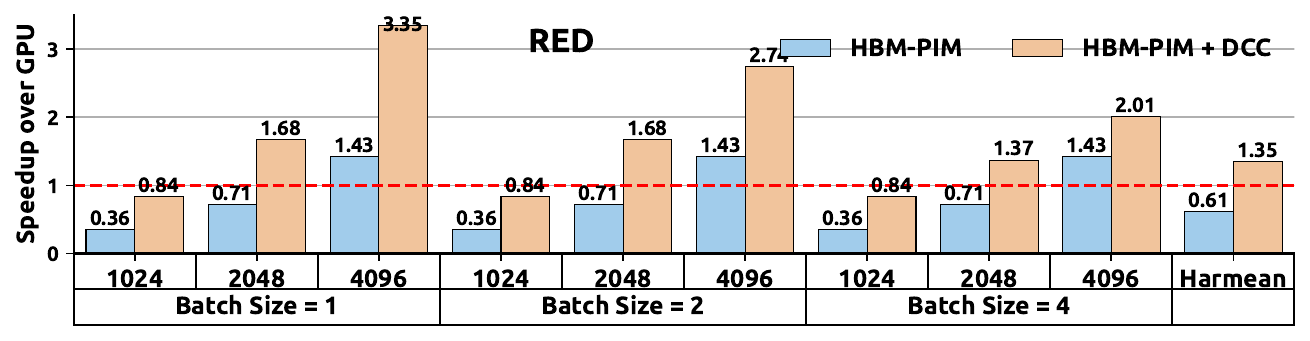}
    \label{fig:speedup_attaccattn_e}
\end{subfigure}

\vspace{-\baselineskip}
\vspace{-0.05cm}

\begin{subfigure}{0.47\textwidth}
    \includegraphics[width=\linewidth]{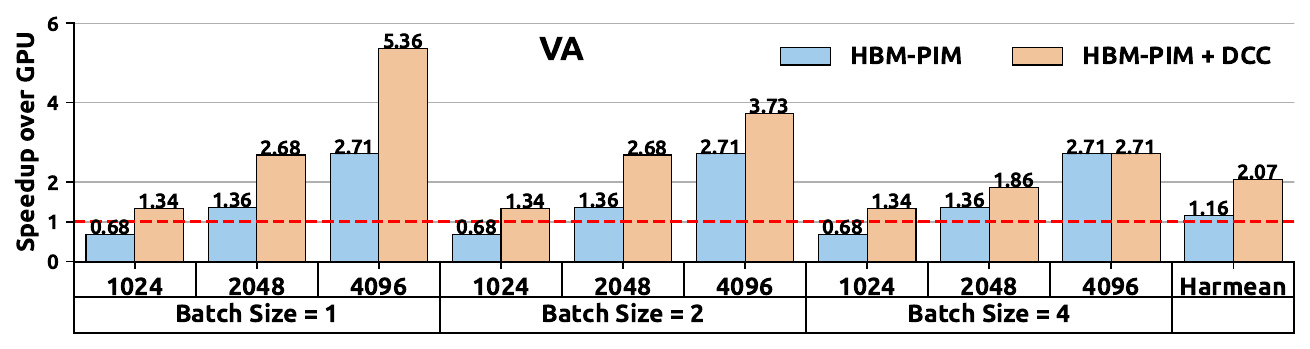}
    \label{fig:speedup_attaccattn_f}
\end{subfigure}

\vspace{-\baselineskip}
\vspace{-0.05cm}

\begin{subfigure}{0.47\textwidth}
    \includegraphics[width=\linewidth]{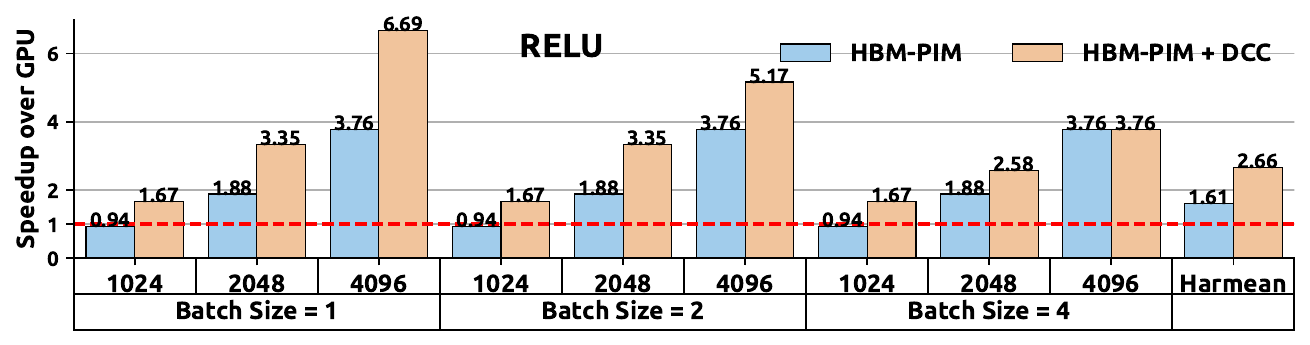}
    \label{fig:speedup_attaccattn_g}
\end{subfigure}

\vspace{-\baselineskip}

\caption{Speedup of HBM-PIM and HBM-PIM+\SysName over GPU for GEMV, RED, VA and RELU, varying the tensor sizes.}
\label{fig:speedup_hbmpim}
\end{figure}

Fig.~\ref{fig:speedup_hbmpim} shows the speedup of HBM-PIM and HBM-PIM+\SysName over  GPU  in various ML kernels, when varying the batch size and tensor sizes. 
We make four key observations. First, \SysName significantly improves HBM-PIM performance by \gemvAvgHbmpimBaselinespeedup$\times$, \redAvgHbmpimBaselinespeedup$\times$, \vaAvgHbmpimBaselinespeedup$\times$, and \reluAvgHbmpimBaselinespeedup$\times$ for GEMV, RED, VA, and RELU on average, respectively, enabling HBM-PIM to further outperform GPU by \gemvAvgHbmpimGpuspeedup$\times$, \redAvgHbmpimGpuspeedup$\times$, \vaAvgHbmpimGpuspeedup$\times$, and \reluAvgHbmpimGpuspeedup$\times$, respectively.
\revB{Second, in large batch sizes and tensor sizes, AttAcc's  distribution is highly optimized over GPUs. Thus, in both HBM-PIM and AttAcc backends (Figs.~\ref{fig:attacc_speedups}, ~\ref{fig:speedup_hbmpim}), \SysName exploits limited optimization opportunities, and enables smaller speedups over GPUs.}
Third,  when using the same batch and tensor size configurations for both HBM-PIM and AttAcc in GEMV and RED, \SysName achieves \gemvAvgHbmpimBaselinespeedup$\times$ and \redAvgHbmpimBaselinespeedup$\times$ average speedup over HBM-PIM, respectively, and \gemvAvgAttaccBaselinespeedup$\times$ and \redAvgAttaccBaselinespeedup$\times$ average speedup over AttAcc, respectively. \SysName provides greater performance improvements on HBM-PIM than on AttAcc, because HBM-PIM is less optimized in hardware than AttAcc for GEMV, i.e., HBM-PIM includes general SIMD units, while AttAcc has speciliazed GEMV units.
Fourth, for RED with tensor sizes 1024 and 2048, HBM-PIM underperforms GPU by \redSmallAvgHbmpimGpuSlowdown$\times$ averaged across all batch sizes, because the combination of small tensor sizes and HBM-PIM's fixed tiling scheme results in poor SIMD utilization: the per-core tensor partition size does not align with the 16-way SIMD instruction width. However, \SysName improves their performance by \redSmallAvgHbmpimBaselineSpeedup$\times$ on average, enabling HBM-PIM to outperform GPU at tensor size 2048.
Overall, \SysName demonstrates robustness across multiple PIM backends through its multi-layer PIM abstraction, providing consistent performance improvements across diverse ML workloads on both HBM-PIM and AttAcc.

\newcommand{\batchOneAvgHbmpimBaselineSpeedup}{1.33}
\newcommand{\batchEightAvgHbmpimBaselineSpeedup}{\revC{1.58}}
\newcommand{\batchOneAvgAttaccBaselineSpeedup}{1.14}
\newcommand{\batchEightAvgAttaccBaselineSpeedup}{\revC{1.31}}

\begin{figure}[ht]
    \centering
    \includegraphics[width=0.99\linewidth, trim={0 0.33cm 0 0}, clip]{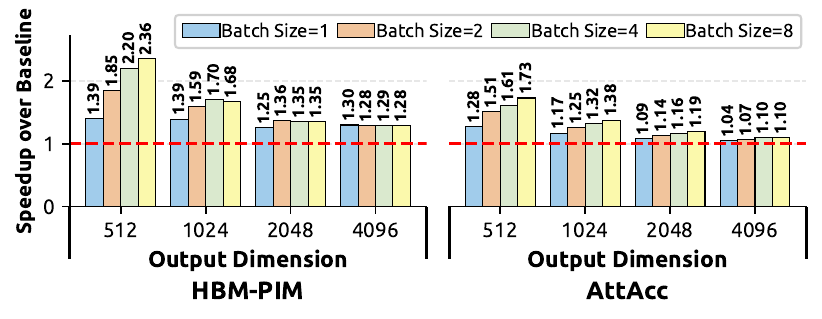}
    \caption{Speedup of \SysName over HBM-PIM (left) and AttAcc (right), when increasing the batch size in the GEMV kernel.}
    \label{fig:sensitivity}
\end{figure}

Fig.~\ref{fig:sensitivity} shows the speedup of \SysName over HBM-PIM and AttAcc, when increasing the batch size in GEMV, evaluating in an attention layer with head count 32, input size 128 and various output sizes. \revC{In both PIM backends, \SysName improves performance over the baseline as the batch size increases. 
On average, \SysName performance gains increase with batch size, from \batchOneAvgHbmpimBaselineSpeedup$\times$ to \batchEightAvgHbmpimBaselineSpeedup$\times$ on HBM-PIM, and from \batchOneAvgAttaccBaselineSpeedup$\times$ to \batchEightAvgAttaccBaselineSpeedup$\times$ on AttAcc, as batch size grows from 1 to 8.
In large output dimension and batch sizes, AttAcc’s GEMV data distribution strategy achieves high performance, leaving limited room for further optimization. In those workloads, \SysName is still able to \emph{automatically} match and slightly improve AttAcc’s hand-optimized performance without programmer intervention, demonstrating that \SysName can provide both high performance and high programming ease.
}

\begin{figure*}
    \centering
    \includegraphics[width=\linewidth]{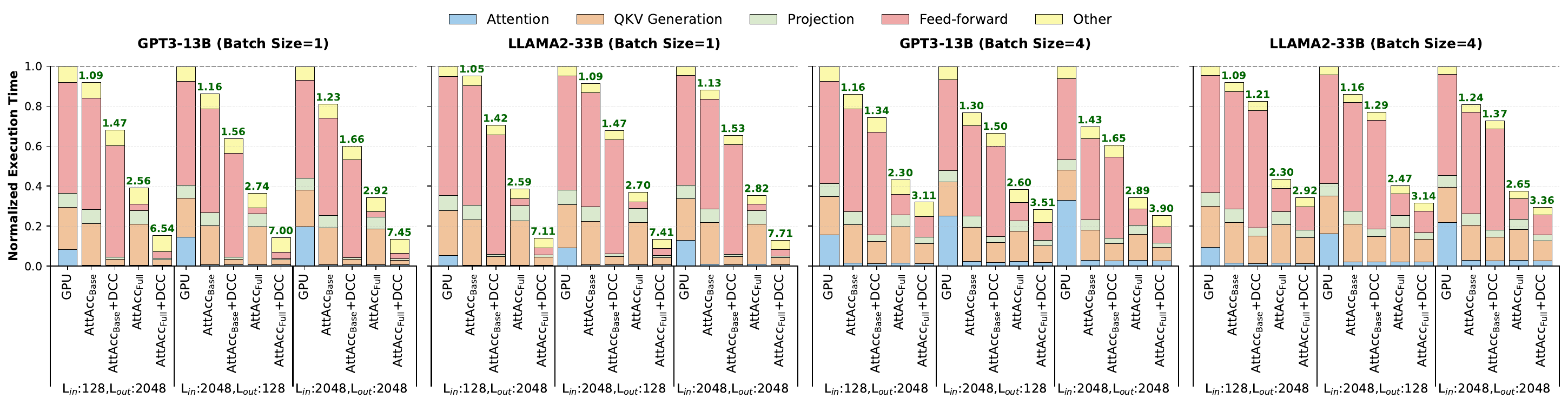}
    \caption{Normalized  time breakdown for GPT3-13B and LLaMA2-33B models for various input, output token and batch sizes. }
    \label{fig:inference}
\end{figure*}


\subsection{End-to-End LLM Inference}
\label{sec:inference}

Fig.~\ref{fig:inference} presents the normalized execution time breakdown of AttAcc and AttAcc+\SysName over GPU for the main computational phases in inference of two state-of-the-art models, while varying the input and output token sizes and batch sizes. We evaluate two variants: AttAcc$_{Base}$ runs only attention layers on PIM, and Attacc$_{Full}$ runs attention layers and a portion of the Feed-forward on the PIM side (See~\cite{park2024attacc}), while the largest portion of Feed-forward runs on GPU. For both AttAcc variants, QKV generation, projection and Other run on GPU.  However, with \SysName's optimizations, we enable QKV generation and projection to be also executed on the PIM side, achieving performance benefits over running them on GPU. In LLM inference, \SysName initializes the kernels with few random requests, then generates tiling drafts on-the-fly for new tensor sizes; this generation time is included in our measurements. The numbers in the top of each bar show speedup over GPU.




\newcommand{\baseAvgAttaccGpuspeedup}{\revC{1.44}}
\newcommand{\fullAvgAttaccGpuspeedup}{\inferenceAvgattaccgpuspeedup}
\newcommand{\baseAvgAttaccBaselinespeedup}{\revC{1.23}}
\newcommand{\fullAvgAttaccBaselinespeedup}{\inferenceAvgattaccBaselinespeedup}
\newcommand{\gptAttnAvgAttaccGpuspeedup}{1.12}
\newcommand{\llamaAttnAvgAttaccGpuspeedup}{1.16}
\newcommand{\qkvAvgAttaccGpuspeedup}{\revC{2.58}}
\newcommand{\prjFCAvgAttaccGpuspeedup}{\revC{2.91}}

We make three key observations.
First, \SysName provides high performance benefits on both Attacc$_{Base}$ and Attacc$_{Full}$ by \pyang{\baseAvgAttaccBaselinespeedup$\times$} and \fullAvgAttaccBaselinespeedup$\times$ on average, respectively, improving performance over GPU by \pyang{\baseAvgAttaccGpuspeedup$\times$} and \fullAvgAttaccGpuspeedup$\times$ on average, respectively.
\SysName significantly strengthens AttAcc, enabling it to substantially outperform GPU across different models and token sizes.
Second, \SysName improves performance on Attention layers by on average \gptAttnAvgAttaccGpuspeedup$\times$ and \llamaAttnAvgAttaccGpuspeedup$\times$ over AttAcc for GPT3-13B and LAMMA2-33B model, respectively. AttAcc uses a fixed tiling strategy across all token counts and batch sizes, while \SysName adapts tiling strategies to different workload configurations.
Third, in QKV generation and projection layers, \SysName achieves \qkvAvgAttaccGpuspeedup$\times$ and \prjFCAvgAttaccGpuspeedup$\times$ speedup over GPU, respectively.
In both AttAcc variants, these layers run on GPU, because AttAcc's fixed tiling strategy 
underperforms GPU by 1.25$\times$ in these layers. In contrast, \SysName comprehensively explores the data-compute co-optimization space to identify optimal tiling configurations, enabling these layers to run efficiently on PIM and significantly outperform GPU.
Overall, we conclude that \SysName provides significant performance benefits in various state-of-the-art LLMs with different token count and batch sizes. These results demonstrate that \SysName can serve as a practical and effective compiler for heterogeneous ML acceleration on high-performance processors (e.g., GPUs) and PIM backends. 

\newcommand{\compAllAvgAttaccspeedup}{\revC{1.58}}
\newcommand{\dataAllAvgAttaccspeedup}{\revC{1.65}}

\newcommand{\compGroupAllAvgAttaccspeedup}{1.18}
\newcommand{\dataGroupAllAvgAttaccspeedup}{\revC{1.67}}

\newcommand{\largeBaseAvgAttaccspeedup}{1.59}
\newcommand{\largeFullAvgAttaccspeedup}{1.67}
\newcommand{\largeQVKAvgAttaccspeedup}{2.21}

\subsection{Large-Scale Experiments}\label{eval:large-scale}

\begin{figure}[!htb]
    \centering
    \includegraphics[width=\linewidth]{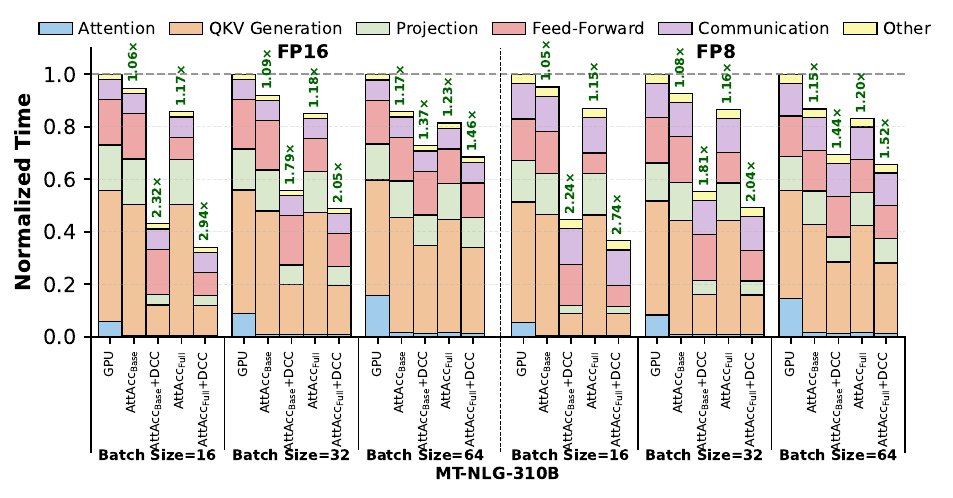}
    \caption{\revF{Normalized time breakdown with MT-NLG-310B model for various batch sizes and numerical precisions using 8 A100 GPUs (DGX-class) with 5 PIM devices each GPU.}}
    \label{fig:large-scale}
\end{figure}

Fig.~~\ref{fig:large-scale} \revF{shows the normalized time breakdown of AttAcc variations and AttAcc+\SysName schemes over GPU baseline using MT-NLG-310B model,  128 input tokens, 2048 output tokens, FP16 and FP8 numerical precisions, and large batch sizes (up to 64). We use 8 A100 GPUs (DGX-class) with 5 HBM-type PIM devices each GPU.  \SysName 
significantly improves performance over Attacc$_{Base}$ and Attacc$_{Full}$ by \largeBaseAvgAttaccspeedup$\times$ and \largeFullAvgAttaccspeedup$\times$, respectively.
In large batch sizes, AttAcc’s GEMV data distribution (QKV generation and projection) already enables high performance. \SysName improves it further by \largeQVKAvgAttaccspeedup$\times$, providing both high performance and high programming ease.
\SysName is highly efficient in larger batch sizes, larger LLMs, large multi-GPU systems and using different numerical precisions,  providing robustness even in large-scale inference scenarios.
}

\subsection{ML Kernel Time Breakdown}\label{eval:kernel-breakdown}

\begin{figure}[!htb]
\captionsetup[subfigure]{labelformat=empty, skip=0pt}
\centering



\includegraphics[width=\linewidth]
{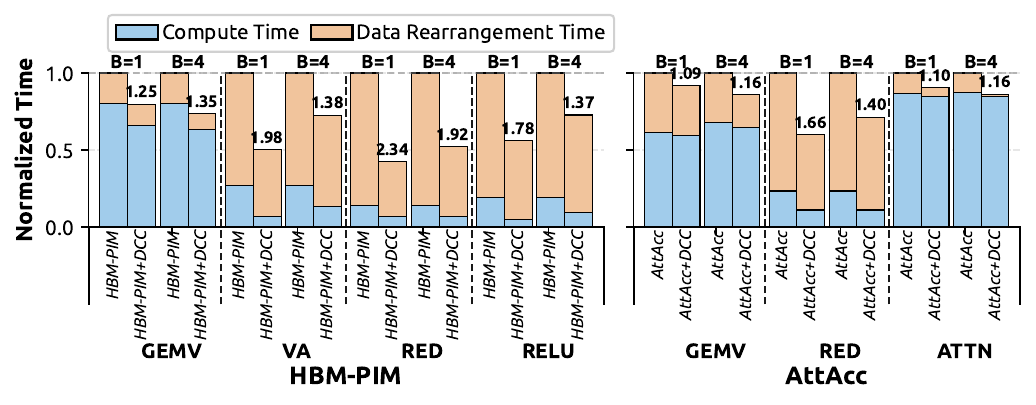}



\caption{Normalized time breakdown of compute and data rearrangement time in various ML workloads and backends.}
\label{fig:execution_breakdown_all}
\end{figure}

Fig.~\ref{fig:execution_breakdown_all} shows the normalized time breakdown split into time spent on \emph{compute} and \emph{data rearrangements} for various ML kernels using two batch sizes, and two different backends with and without \SysName. The numbers above bars show the speedup provided by \SysName. The execution time is normalized to that of the respective PIM backend with its default implementation.


We draw two findings. First, \SysName co-optimizes both compute time and data rearrangement time, providing \compAllAvgAttaccspeedup$\times$ and \dataAllAvgAttaccspeedup$\times$ speedup, respectively. 
\SysName comprehensively explores a broader joint optimization space, 
and identifies  \revS{highly efficient} balance between data movement and computation costs.
Second, \SysName provides larger  benefits in ML kernels, where data rearrangement dominates execution time. The VA, RED, and RELU kernels  exhibit substantial data rearrangement overheads, and \SysName accelerates them by \dataGroupAllAvgAttaccspeedup$\times$ on average, while it accelerates the compute-heavy GEMV and ATTN kernels  by \compGroupAllAvgAttaccspeedup$\times$. These results indicate  that 
\SysName significantly alleviates data rearrangement bottlenecks in PIM, while also effectively integrating compute-specific optimizations.
\newcommand{\energyAllAvgAttaccspeedup}{1.24}

\subsection{Energy Efficiency Analysis}\label{eval:energy}

\begin{figure}[!htb]
    \centering
    \includegraphics[width=\linewidth]{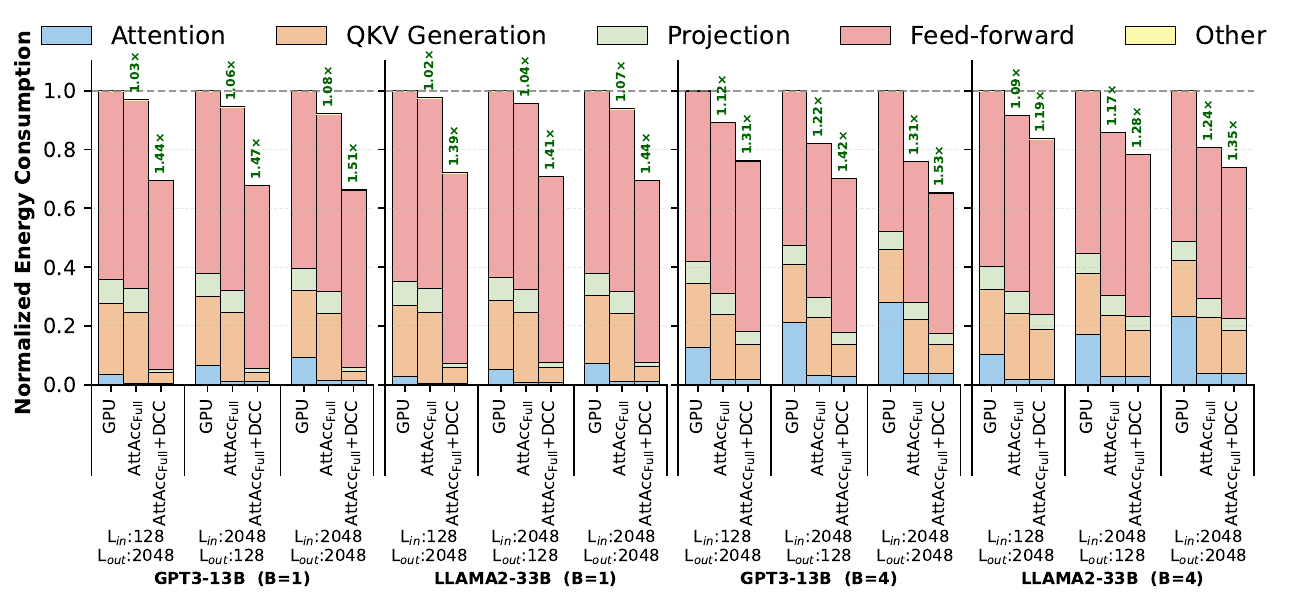}
    \caption{\revB{Normalized energy breakdown in GPT3-13B and LLaMA2-33B comparing AttAcc$_{Full}$ PIM system over GPU. }}
    \label{fig:energy}
\end{figure}

Fig.~\ref{fig:energy} \revB{presents \SysName's in energy efficiency  as kernel breakdown using AttAcc$_{Full}$ system  compared to an A100 GPU in end-to-end inference of various LLMs, input token, output token, and batch sizes.
We follow the methodology of the AttAcc paper~\cite{park2024attacc}. 
AttAcc's energy simulation synthesizes all arithmetic units using Synopsys Design Compiler with a 7nm predictive PDK (ASAP7~\cite{clark2016asap7}), models SRAM-based buffers using FinCACTI~\cite{ShafaeiFinCACTI}, and scales all components to the appropriate DRAM process technology node. Energy consumption inside the DRAM die is calculated using datapath lengths from HBM chip microphotographs~\cite{Ryu2023HBM3,Park2022HBM3}. 
\SysName improves energy efficiency over AttAcc$_{Full}$ on average by \energyAllAvgAttaccspeedup$\times$. 
\SysName's performance benefits translates to energy efficiency gains.}

\newcommand{\pruneSearchSpaceReduction}{9.01}
\newcommand{\pruneCompilationTimeReduction}{5.97}
\newcommand{\predictorCompilationTimeReduction}{3.81}

\subsection{Compilation Time Evaluation}
We \revF{evaluate \SysName's compilation time, and the effectiveness of its draft pruning scheme and its prediction model.
Table~\ref{tab:kernel_prune_perf} shows the \SysName search space (\#candidate drafts) and compilation time, when disabling versus enabling draft pruning for a single workload on AttAcc. We use 128 input tokens, 128 output tokens, 32 heads for ATTN or GEMV, and we vary the PIM devices per GPU and batch sizes. }

We \revF{draw three findings. First, \SysName compile time is low (a few seconds), occurring \textbf{\emph{only once}} offline before inference. In real deployment scenarios with billions of  inference requests on the same model and system configuration, the amortized compilation cost is negligible.
Notably, compilation time of a single kernel does not significantly increase, when considering more tensor shapes: ATTN with 1 tensor shape takes 2.96s and with 272 different tensor shapes takes 4.74s (Table~\ref{tab:predictor_acc}). Increasing tensor shapes during compilation primarily increases GPU utilization rather than compilation time. 
Second, the search space primarily grows with system scale, e.g., adding more PIM devices per GPU, as larger systems introduce more candidate data partitioning strategies. Increasing batch sizes (or tensor shapes that we also tested) with a fixed number of PIM devices/cores only slightly increases the search space (the number of candidate partitioning strategies across cores remains similar).
Third, our pruner significantly reduces the search space by on average \pruneSearchSpaceReduction $\times$ and compilation time by \pruneCompilationTimeReduction $\times$, with no impact on kernel execution performance. \SysName's pruning rules eliminate largely inefficient configurations, ensuring that \SysName reliably finds near-optimal schedules.}

\begin{table}[!htb]
  \centering
  \resizebox{1\linewidth}{!}{
  \revF{
  \begin{tabular}{|l|c|c|c|c|c|c|c|}
    \hline
    \rowcolor{teal!16} \textbf{Kernel} & 
    \textbf{\makecell{PIM devices\\per GPU}} & 
    \textbf{\makecell{batch\\size}} & 
    \textbf{\makecell{Number of\\GPUs}} & 

    \textbf{\makecell{\#drafts\\w/o pruning}} & 
    \textbf{\makecell{\#drafts\\w pruning}} & 
    \textbf{\makecell{compilation time\\w/o pruning (s)}} & 
    \textbf{\makecell{compilation time\\w pruning (s)}} \\
    \hline 
    ATTN  & 5 & 4 & 1  & 32186 & 4274 & 9.57   & 2.96  \\ 
    ATTN  & 5 & 64 & 8  & 33271 & 4868 & 16.20  & 4.10  \\
    ATTN  & 10 & 4 & 1  & 79952 & 6200 & 29.90  & 4.22  \\
    ATTN  & 10 & 64 & 8 & 83708 & 7319 & 54.51  & 6.32  \\
    GEMV  & 5 & 4 & 1  & 32186 & 4274 & 19.75  & 3.75  \\
    GEMV  & 5 & 64 & 8 & 33271 & 4868 & 36.86  & 5.54  \\
    GEMV  & 10 & 4 & 1  & 79952 & 6200 & 57.66  & 5.46  \\
    GEMV  & 10 & 64 & 8 & 83708 & 7319 & 109.92 & 9.30  \\
    \hline
  \end{tabular}
  }
  }
  \caption{\revF{\SysName search space and compilation time, when disabling versus enabling the pruning for single kernel workload.}}
  \label{tab:kernel_prune_perf}
\end{table}

We run ML kernels on HBM-PIM and AttAcc with hundreds of different tensor and batch sizes per kernel, representative configurations of real ML models, \revB{and all  kernel configurations of all evaluated LLMs. \revS{We use \SysName (pruning is disabled) to exhaustively generate all possible tiling drafts, execute all of them (the whole search space) on the target PIM backend (not using our predictor), and identify the true optimum draft.}
Table~\ref{tab:predictor_acc} compares the \SysName predictor's selections against the true optimum drafts. The \textit{Total} column shows the number of tensor shapes and batch sizes configurations tested per kernel. The \textit{\#Best} column shows how many cases our predictor correctly identified the true optimum draft.
The \textit{Compilation Time w/o Predictor} is the \SysName compile time when running the workloads on an A100 GPU.
The \textit{Compilation Time w Predictor} is the \SysName compile time using our predictor.
The \textit{Suboptimal Kernel Performance} and \textit{Suboptimal Total Performance} columns show the average 
performance degradation at kernel and end-to-end inference level, respectively, for the cases where the predictor selects a suboptimal schedule.

\begin{table}[!htb]
\centering
\resizebox{0.999\linewidth}{!}{
\begin{tabular}{|l|c|c|c|c|c|c|}
\hline
\rowcolor{teal!16} \textbf{Backend+\revB{Workload}} & 
\textbf{Total} & 
\textbf{\#Best } & 
\textbf{\makecell{\revB{Compilation Time}\\ \revB{w/o Predictor (s)}}} & 
\textbf{\makecell{\revB{Compilation Time}\\ \revB{w Predictor (s)}}}  &
\textbf{\makecell{\revB{Suboptimal Kernel}\\ \revB{Performance}}} & 
\textbf{\makecell{\revB{Suboptimal }\\ \revB{Total Performance}}}\\
\hline
AttAcc+ATTN & 272 & 203 & \revB{24.80}  & \revB{4.74}  & 97.04\% (69 cases) & \revB{-}\\
AttAcc+GEMV & 304 & 290 & \revB{60.70}  & \revB{11.21}  & 97.01\% (14 cases) & \revB{-}\\
AttAcc+RED  & 167 & 155 & \revB{7.60}   & \revB{2.22}  & 94.01\% (12 cases) & \revB{-}\\
\revB{Attacc+GPT3} & \revB{502} & \revB{489} & \revB{11.94} & \revB{4.84} & \revB{97.94\% (13 cases)} & \revB{99.89\%}\\
\revB{Attacc+LLAMA2} & \revB{502} & \revB{488} & \revB{16.13}  & \revB{5.49}   & \revB{98.83\% (14 cases)} & \revB{99.57\%}\\
\revB{Attacc+MT-NLG} & \revB{251} & \revB{245} & \revB{56.53}  & \revB{11.22}  & \revB{99.07\% (6 cases)} & \revB{99.80\%}\\
HBM-PIM+GEMV & 304 & 265 & \revB{116.07} & \revB{19.50}  & 95.84\% (39 cases) & \revB{-}\\
HBM-PIM+RED  & 167 & 154 & \revB{8.09}   & \revB{2.29}   & 96.43\% (13 cases) & \revB{-}\\
HBM-PIM+VA   & 167 & 158 & \revB{14.21}  & \revB{3.24}  & 94.94\% (9 cases) & \revB{-}\\
HBM-PIM+RELU & 167 & 157 & \revB{5.76}   & \revB{1.94}  & 94.48\% (10 cases) & \revB{-}\\

\hline
\end{tabular}
}
\caption{Prediction accuracy, \revS{compilation time and suboptimal performance using \SysName} across various ML workloads.}
\label{tab:predictor_acc}
\end{table}

The \SysName's predictor achieves on average 89.28\%  and 97.37\% accuracy on single kernels and LLMs, respectively, in correctly identifying the true optimum schedules. When suboptimal, the predictor's selections achieve 96.56\% of true optimal performance on average across different workloads and backends. In AttAcc+ATTN, it has relatively lower accuracy (74.63\%), since ATTN fuses GEMV and softmax as a single kernel. However, it still provides high performance (97.4\% of true optimum). Even in large LLM inference, \SysName's predictor rarely finds sub-optimal schedules, having only $\sim$0.2\% total performance slowdown over when using the true optimum schedules. 
Moreover, our predictor significantly accelerates compilation time by \predictorCompilationTimeReduction $\times$ averaged across all workloads, LLMs, and backends.
Overall, our results show that even when our predictor fails to identify the true optimum, the selected schedule has negligible performance degradation, while significantly accelerating the compilation time. }



\subsection{Dynamic Prediction Evaluation}

\begin{figure}[!htb]
    \centering
    \includegraphics[width=0.856\linewidth]{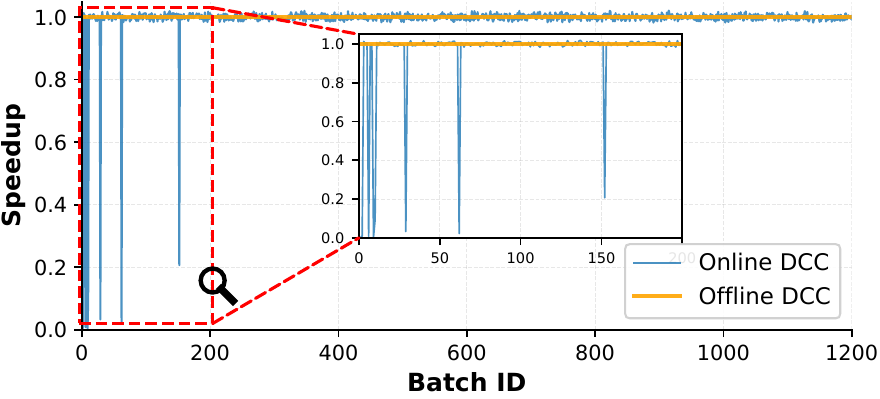}
    \caption{\revF{Online vs offline \SysName on MT-NLG 310B inference.}}
    \label{fig:online-inf}
\end{figure}

We \revF{evaluate \SysName in an online draft selection scenario with unseen configurations: for given ML kernel workloads \SysName's lookup table has not recorded the best-performing schedules. 
\revS{We execute 2000 batches (batch size 64) of synthetic inference queries on MT-NLG-310B, with input and output token sizes randomly generated from the range [128, 2048], covering representative sequence lengths of real-world deployments.} Fig.~\ref{fig:online-inf} compares \emph{online \SysName} with unseen tensor shapes against \emph{offline \SysName}, where all tensor shapes have been previously seen and are recorded in the lookup table.
Online \SysName incurs performance penalty only during the first $\sim$160 queries, where unseen tensor shapes trigger additional compilation. After this short warm-up period, \SysName's lookup table covers all encountered tensor shapes and online \SysName converges to the same performance as offline \SysName. \revS{\SysName's online overhead is short-lived, requiring only a few hundred warm-up queries, and is fully amortized in real-world LLM deployments where the number of requests (typically millions of LLM queries) is orders of magnitude larger than \SysName's warm-up period.
}}

\vspace{-6pt}
\section{Related Work}
\vspace{-6pt}
To our knowledge, this is the first work to (i) consider data rearrangement strategies and their associated costs during ML kernel tuning for xPU-integrated PIM devices, and (ii) propose a compiler that jointly optimizes them  with compute code. 

\noindent\textbf{Compiler Support for PIM.}
Prior works~\cite{pimflow_cgo2023,khan2024cinm,pimdl_asplos2024, simplepim_pact2023} design compilation tools for PIM systems, but lack systematic optimization and auto-tuning capabilities. They target a single ML kernel or 
a single PIM backend.
ATiM~\cite{atim_isca2025} is a search-based tensor compiler for UPMEM PIM. However, UPMEM's DDR4-based architecture targets CPU memory channels, thus preventing GPU-PIM co-executions for ML models. 
ATiM \revS{employs} \emph{compute-centric} tuning, generating data rearrangement strategies \textbf{in isolation} from compute code generation.
Such compute-centric process yields sub-optimal performance (§\ref{sec:motivation-datacentric}).
Instead, \SysName has a \emph{unified data-centric} approach that \textbf{co-optimizes} \cg{compute-data to minimize end-to-end time}.

\noindent\textbf{Compiler Support for Processing-Using-Memory (PUM).} OptiPIM~\cite{optipim_asplos2025}, MVE~\cite{khadem2025multi}, and TCCIM~\cite{drebes2020tc} \revS{propose} compilers for PUM devices, \revS{which} are analog-based and have higher hardware design complexity than PIM devices (digital-based). 
\noindent\textbf{PIM Architectures and Accelerators}. 
UPMEM PIM~\cite{upmem,Gomez2022Benchmarking} DDR4-based device for CPUs \cg{lacks} a complete 32$\times$32-bit integer multiplier and 
floating-point units, making it unsuitable for our target ML workloads.
Samsung HBM-PIM~\cite{lee2021hardware} and SK Hynix GDDR6-AiM~\cite{lee20221ynm,He2020Newton} are 3D memory devices with  floating-point units, can be integrated with 
GPUs, 
and have been validated in real systems. 
Numerous research works~\cite{park2024attacc,kim2025cost,li2025blockpim,liu2025mcpal,chen2025attenpim,he2025papi,li2024specpim,pimgpt_natcomm2024,lee2025paise,kim2025pimba,quinn2025longsight} enhance near-bank PIM devices to support critical ML primitives (e.g., GEMV, ReLU, Softmax), can be integrated with xPUs, and have floating-point capabilities. 
\SysName can be directly used on them to automate and accelerate ML kernels.
Near-rank PIM systems~\cite{rhee2025hpu,kwon2019tensordimm,liu2025make,gu2025pim} place cores at the buffer chip of the DIMM with access to all banks. Despite different core placement,  \SysName can effectively support near-rank PIM designs for ML.
Recent works propose hybrid near-rank and near-bank designs~\cite{jang2025accelerating, quinn2025drex, kim2024darwin} and NPU-integrated PIM devices~\cite{he2025lp,han2025near,wu2025pimoe,ianus_asplos2024,heo2024neupims,Ruiyang2025HEAT}. \SysName can be easily extended to support them thanks to its multi-layer abstraction.
Finally, prior works~\cite{ahn2015scalable,dai2018graphh, kang2025sparsepim, sparsep_pomacs2022, kwon2019tensordimm,lee2024pim,zhou2022gnnear,tian2022GNMP,yun2023grande,Chen2023MetaNMP,Saed2025RayN, yan2021copim} design application-specific PIM devices for graph analytics, sparse kernel, or data retrieval. While \SysName may benefit some kernels in these devices, we \revS{primarily target ML kernels and models}.


\noindent\textbf{Software for PIM Systems}. 
Prior works~\cite{Gomez2022Benchmarking,giannoula2022sparsep,Diab2023Framework,gomez2023evaluating,Lim2023Design,Item2023TransPimLib,Das2022Implementation,Jibril2024Aggregation,chen2023simplepim,rhyner2024analysis, giannoula2022towards, pid_join_upmem2023,pimpam_sigmod2024,nider2021case,kang2025pim,kim2025no,kong2025pimbeam,giannoula2024pygim,giannoula2024accelerating,barkhordar2025alpha} propose libraries, frameworks, and benchmark suites for the UPMEM PIM spanning  linear algebra, graph processing, image processing, machine learning, databases, and concurrent data structure domains. UPMEM PIM is primarily designed for CPUs 
and has limited hardware multiplication support. 
\SysName is designed for PIM devices that efficiently support ML kernels.

\noindent\textbf{Communication and System Integration for PIM}. 
Prior works~\cite{Seunghyuk2025ComPASS,pid_comm_2024,pim_mmu_micro2024,um_pim_isca2024,vpim_isca2024,pimcare_ics2025,giannoula2021syncron,pim_simulator_2021, hyun2024pathfinding,heo2023primo,Kim2025PIMCCA} design efficient  data transfers, memory management, synchronization, virtualization of PIM, and simulation tools.
PIMCARE~\cite{pimcare_ics2025} is a compiler-assisted scheduler to allocate the correct amount of PIM devices for an application.
\revA{UniNDP~\cite{Xie2025UniNDP} is a simulation and compilation framework for different PIM architecture types (near-bank, near-device, near-rank, near-channel) to guide programmers how to develop their kernels. To run a kernel on a target PIM backend, programmers still need to manually implement it using backend-specific low-level programming interface. 
Instead, \SysName 
directly takes kernels written in high-level Python interface (§\ref{sec:mechanism-interface}), and automatically generates optimized low-level code for a PIM backend. Programmers using \SysName require \textbf{no} expertise in the target PIM backend’s programming interface or architecture.
}


\noindent\textbf{ML Compilers for Commodity Systems}.
\revC{
Prior works~\cite{Gupta2025SPLAT,won2023unified,Chen2018TVM,zheng2020ansor,chen2018learning,Nakandala2020Tensor,XLA2020Compiler,Ding2023Hidet,feng2023tensorir,Kjolstad2017TACO,Ye2023SparseTIR,Ahrens2025Finch,Liu2025CROSS,Du2025SRSparse,xing2022bolt,Pouget2025Holistic} propose ML compilers for CPU, FPGA and GPU systems by leveraging their shared memory model and deep cache hierarchies (on-chip caches).} 
\revA{Prometheus~\cite{Pouget2025Holistic} optimizes kernels on FPGAs via loop transformations for cache locality, and overlapping computation with off-chip 
data transfers to main memory.
Instead,  PIM systems have a distributed memory model and shallow cache hierarchy, and may not have hardware support to overlap PIM computation with off-chip data transfers. 
Thus, 
compiler techniques such as cache tiling, loop transformations, and compute-communication overlapping cannot enable high performance on PIM. 
} 
\revC{ML compilers for CPU/GPU/FPGA  cannot be directly used in PIM systems, or would cause high performance costs because they do not optimize data how data is distributed across DRAM banks.}

\vspace{-3pt}
\section{Conclusion}
\vspace{-3pt}

We observe that Host xPU processor and PIM cores require different data layouts, necessitating data rearrangements that pose significant performance challenges in ML kernels. We also find that data rearrangement and compute code optimization are interdependent.
To this end, we design \SysName, a data-centric ML compiler for PIM systems, that \emph{co-optimizes} compute code and data rearrangement strategies in a unified tuning process.
\SysName enables up to \kernelmaxhbmpimgpuspeedup$\times$ speedup over GPU on HBM-PIM and up to \kernelmaxattaccgpuspeedup$\times$ speedup over GPU on AttAcc for ML kernels. \SysName on AttAcc accelerates end-to-end LLM inference by up to \inferenceMaxattaccgpuspeedup$\times$. We hope our work encourages further research on compilation tools  for  ML kernels on PIM systems.

\vspace{-6pt}
\section*{Acknowledgements}
\vspace{-1pt}
We thank \CamOne{the anonymous reviewers of ISCA 2026 for valuable feedback. We thank Qidong Su, Zhanda Zhu and Yaoyao Ding for their suggestions. We thank the SPIN Research Group, the SAFARI Research Group and the EcoSystem Research Group for providing a stimulating intellectual environment.}  


\bibliographystyle{IEEEtran}
\bibliography{refs}

\clearpage
\newpage
\appendix

\section{Artifact Appendix}

\subsection{Abstract}

The Artifact Appendix describes how to reproduce the main results of this paper. It includes the source code of \SysName, benchmark scripts,  and step-by-step instructions for the key evaluation results.  The experiments require a server equipped with a CPU with at least 64 hardware threads, 128GB of main memory, disk space of at least 128GB, and an NVIDIA GPU with an up-to-date NVIDIA driver installed. 
Note that our artifact needs Anaconda 25.6.1+ and CMake 3.16.3 to be installed in the server.
We provide a \texttt{README.md} file that describes the required hardware and software dependencies and provides step-by-step instructions. 
This artifact is used to support our major claims (See Appendix §\ref{sec:claims}), demonstrating \SysName's performance benefits in Figures \ref{fig:attacc_speedups}, \ref{fig:speedup_hbmpim} and \ref{fig:inference}.
We expect the full evaluation pipeline, including the setup and simulations, to take approximately 7-10 days. We also expect that the trace files generated for trace-driven simulations take approximately 90GB disk space.

\subsection{Artifact Check-list (meta-information)}

\begin{itemize}[topsep=0pt,leftmargin=12pt,nosep,partopsep=0pt]
\item {\bf Program: } \textit{DCC\_Artifact}: In this artifact, we evaluate \SysName compiler upon HBM-PIM and AttAcc PIM backends and compare HBM-PIM, HBM-PIM + \SysName, AttAcc, AttAcc + \SysName comparison points against GPU baseline. 
\item {\bf Compilation: } This artifact \textbf{strictly} requires CMake 3.16.3 and GCC/G++ 11.4.
\item {\bf Models: } The workloads for kernel performance and inference performance have been configured to evaluate the GPT3-13B and LLAMA2-33B models.
\item {\bf Run-time environment:} Linux Ubuntu 20.04 (or newer) with Python 3.8, requiring CUDA 12.3 (or newer).
\item {\bf Hardware:} A server equipped with a CPU of at least 64 hardware threads and 128GB of main memory, and an NVIDIA GPU device with a minimum compute capability of 8.0 and at least 8GB GPU memory should be used to validate the results.
\item {\bf Execution:} Trace-driven simulations.
\item {\bf Metrics:} Execution time normalized as relative performance speedup.
\item {\bf Output:} Experimental results
are stored as CSV files, and our scripts generate corresponding figures in
PDF format. The generated figures are similar to the Figures \ref{fig:attacc_speedups}, \ref{fig:speedup_hbmpim}, and \ref{fig:inference} of the main paper.
\item {\bf Experiments:} Kernel performance and end-to-end inference performance. Detailed steps for the reproduction of evaluation experiments are provided in the \texttt{README.md} file.
\item {\bf How much disk space required (approximately)?: } 128 GB. 
\item {\bf How much time is needed to prepare workflow (approximately)?: } 10 minutes (build the code and download Python packages). 
\item {\bf How much time is needed to complete experiments (approximately)?: } 7-10 days (run simulations and generate figures). 
\item {\bf Publicly available?: }Yes.
\item {\bf Code licenses: } MIT. 
\item {\bf Archived: } \url{https://doi.org/10.5281/zenodo.19442321}
\end{itemize}

\subsection{Description}

\subsubsection{How to Access} \label{sec:appendix-source-code}
Download the compressed file DCC\_Artifact.zip from the Zenodo archive \url{https://doi.org/10.5281/zenodo.19442321} or
our GitHub repository at \url{https://github.com/SPIN-Research-Group/DCC}.

\subsubsection{Hardware Dependencies}

The artifact should be tested on a sever with:

\begin{itemize}[topsep=0pt, leftmargin=12pt, partopsep=0pt, parsep=0pt, itemsep=0pt, noitemsep, before=\vspace{-\parskip}]
\item x86-64 CPU with at least 64 hardware threads, 128GB of main memory and 128GB of disk storage.
\item NVIDIA GPU device with a minimum compute capability (SM) of 8.0 and at least 8GB GPU memory.
\end{itemize}


\subsubsection{Software Dependencies}

The artifact requires the following software for installation:

\begin{itemize}[topsep=0pt, leftmargin=12pt, partopsep=0pt, parsep=0pt, itemsep=0pt, noitemsep, before=\vspace{-\parskip}]
\item Ubuntu 20.04 (or newer)
\item Python 3.8
\item GNU compilers (gcc/g++) 11.4.0 (\textbf{strict} requirement)
\item CUDA 12.3 (or newer)
\item CMake 3.16.3 (\textbf{strict} requirement)
\item Anaconda 25.6.1 (or newer)
\item Package scikit-learn 1.3.2 (or newer)
\item Package XGBoost 2.1.4 (or newer)
\end{itemize}

\subsection{Installation}

Download the zip file containing the artifact source code in §\ref{sec:appendix-source-code}.
We provide detailed instructions in the \texttt{README.md} file under the root of source code directory to build the simulator and evaluate \SysName on two PIM backends.

We next summarize the key
steps:
\begin{enumerate}[topsep=0pt,leftmargin=12pt,nosep,partopsep=0pt]
    \item Install Anaconda 25.6.1 (or newer) following the instructions provided in
\url{https://www.anaconda.com/docs/getting-started/main}. You may need to run the following command after installation to enable conda to create a new virtual environment. Please refer to \url{https://www.anaconda.com/docs/getting-started/tos-plugin}.
    \begin{tcolorbox}[
      colback=gray!10, 
      colframe=gray!50, 
      boxrule=0.4pt, 
      left=4pt, right=4pt, top=2pt, bottom=2pt,
      halign=flush left,
      sharp corners
    ]
    \ttfamily
    \$ conda tos accept
    \end{tcolorbox}
    \item Download and install Cmake 3.16.3 (\textbf{strict} requirement) following the instructions provided in
\url{https://cmake.org/download/}. Please note that we have not extensively tested all various CMake versions. We \textbf{recommend} using and building our artifact with Cmake 3.16.3.
    \item Download the source code of DCC\_Artifact  (See §\ref{sec:appendix-source-code}).
    \item Setup the runtime environment for all experiments by running the following bash setup script under the \texttt{root} of source code directory: 
    \begin{tcolorbox}[
      colback=gray!10, 
      colframe=gray!50, 
      boxrule=0.4pt, 
      left=4pt, right=4pt, top=2pt, bottom=2pt,
      halign=flush left,
      sharp corners
    ]
    \ttfamily
    \$ bash setup.sh
    \end{tcolorbox}
\end{enumerate}

\subsection{Experiment Workflow}

The artifact contains three experiments to conduct the evaluation  of the Figures \ref{fig:attacc_speedups}, \ref{fig:speedup_hbmpim}, and \ref{fig:inference} of the main paper. We have a bash script to launch all the simulations of the three evaluation experiments,  collect the raw results and save them under the \texttt{results/} directory. We also include plotting scripts to parse the raw results and generate the figures under  the \texttt{figures/} directory. Next, we provide more details.

\noindent\textbf{Launch Experiments \& Visualize the Results: } We strongly \textbf{recommend} using a server with at least 64 hardware threads and at least 128GB of main memory. The following script (i) launches all the experiments required to reproduce the key results of our paper, (ii) stores the raw results under the \texttt{results/} directory, and (iii) generates the Figures \ref{fig:attacc_speedups} and \ref{fig:speedup_hbmpim}, \ref{fig:inference}  under the \texttt{figures/} directory:
\begin{tcolorbox}[colback=gray!10, colframe=gray!50, boxrule=0.4pt, left=4pt, right=4pt, top=2pt, bottom=2pt]
\texttt{\$ bash run\_experiments.sh}
\end{tcolorbox}

\noindent\textbf{Relaunch Failed Experiments (if any): }If there are any failed simulations (the plotting scripts may also fail to visualize the figures), re-run the main bash script:
\begin{tcolorbox}[colback=gray!10, colframe=gray!50, boxrule=0.4pt, left=4pt, right=4pt, top=2pt, bottom=2pt]
\texttt{\$ bash run\_experiments.sh}
\end{tcolorbox}

\subsection{Evaluation and Expected Result}\label{sec:claims}

\textbf{Major Claims}. For each of the three experiments and for the same workload configuration, we expect the reproduced results to be similar to those reported in the paper within ±4\% due to randomness in the \SysName's predictor offline training process. Specifically, \SysName's predictor is trained offline by selecting a small set of \emph{random} workload configurations (input token sizes, output token sizes, batch sizes etc.). We will focus on the range of speedup values in each evaluation experiment to verify the reproduction of the key results. We next clarify our major claims:

\begin{enumerate}[topsep=0pt, leftmargin=12pt, partopsep=0pt, parsep=0pt, itemsep=0pt, noitemsep, before=\vspace{-\parskip}]
\item Figure~\ref{fig:attacc_speedups}: AttAcc + \SysName achieves kernel performance speedups for  the workload configurations of the ATTN, GEMV and RED kernels in the range of 9.22$\times$--13.70$\times$, 6.55$\times$--10.18$\times$, and 0.95$\times$--4.99$\times$, respectively, over GPU, and in the range of 1.01$\times$--1.64$\times$, 1.00$\times$--1.80$\times$, and 1.02$\times$--1.73$\times$, respectively, over AttAcc. 

\item Figure~\ref{fig:speedup_hbmpim}: HBM-PIM + \SysName achieves kernel performance speedups  for the  workload configurations of the GEMV, RED, VA, and RELU kernels in the range of 4.63$\times$--7.99$\times$, 0.80$\times$--3.48$\times$, 1.29$\times$--5.58$\times$, and 1.60$\times$--6.96$\times$, respectively, over GPU, and in the range of 1.18$\times$--2.45$\times$, 1.35$\times$--2.44$\times$, 1.35$\times$--2.43$\times$, and 1.34$\times$--2.42$\times$, respectively, over HBM-PIM. 

\item Figure~\ref{fig:inference}: Attacc$_{Base}$ + \SysName achieves end-to-end performance speedups for the workload configurations of the two LLMs in the range of 1.15$\times$--1.73$\times$ over GPU, and in the range of 1.06$\times$--1.41$\times$ over  Attacc$_{Base}$.  Attacc$_{Full}$ + \SysName achieves end-to-end performance speedups in the range of 2.77$\times$--8.02$\times$ over GPU, and in the range of 1.20$\times$--2.85$\times$ over  Attacc$_{Full}$. 
\end{enumerate}

\subsection{Experiment Customization}

In kernel performance experiments,  users can customize the workload configuration, including batch size, input and output token sizes and  number of heads by adding following function call at the end of the \texttt{src/test\_configs.py} script, where the \texttt{name} parameter should be set to \texttt{"ATTN"}, \texttt{"GEMV"}, \texttt{"RED"}, \texttt{"VA"}, or \texttt{"RELU"} depending on the target ML kernel: 
\begin{tcolorbox}[colback=gray!10, colframe=gray!50, boxrule=0.4pt, left=4pt, right=4pt, top=2pt, bottom=2pt]
\texttt{custom\_kernel\_sizes(name, batch, input, output, nhead)}
\end{tcolorbox}

For LLM inference performance experiments, users can customize the workload configuration, including batch size, input and output token sizes, by first adding following function call at the end of  the \texttt{src/test\_configs.py} script: 
\begin{tcolorbox}[colback=gray!10, colframe=gray!50, boxrule=0.4pt, left=4pt, right=4pt, top=2pt, bottom=2pt]
\texttt{custom\_inference\_sizes(batch, input, output)}
\end{tcolorbox}
\noindent and then by adding following command (in one line) at the end of the  \texttt{run\_experiemtns.sh} script:
\begin{tcolorbox}[colback=gray!10, colframe=gray!50, boxrule=0.4pt, left=4pt, right=4pt, top=2pt, bottom=2pt]
\texttt{conda run -n dcc --live-stream python3 run\_inference.py --batch=batch --lin=input --lout=output}
\end{tcolorbox}

Note that the plotting scripts do not support custom experiment configurations. To avoid errors, please comment out the last two lines in the \texttt{run\_experiments.sh} script, and update the corresponding plotting script accordingly to match the custom configurations.

\end{document}